\documentclass[longauth]{aa}
\usepackage[switch]{lineno}
\modulolinenumbers[5]

\usepackage[T1]{fontenc}
\usepackage{ae,aecompl}
\usepackage{natbib,twoopt}
\usepackage{booktabs}
\usepackage{ulem}
\usepackage{placeins}
\usepackage{caption}
\usepackage{booktabs}
\usepackage{makecell} 

\usepackage{graphicx}   % Including figure files
\usepackage{amsmath}    % Advanced maths commands
\usepackage{amssymb}    % Extra maths symbols
\usepackage{gensymb}    % For the \degree symbol
\usepackage{multirow}
\usepackage{textcomp} % For prime and other text symbols
\usepackage{mathtools}
\usepackage{IEEEtrantools}
\usepackage{float}
\usepackage{rotating}
\usepackage{bm}
\usepackage{adjustbox}             % To adjust the width of tables that are a bit too large
\usepackage{subcaption}            % For multiple plots within one figure environment
\usepackage{threeparttable}
\usepackage{txfonts} % must be after \usepackage{amsmath} or you will get errors
\usepackage{footmisc}
\usepackage{hyperref}
\hypersetup{
    colorlinks=true,
    linkcolor=blue,
    filecolor=blue,
    urlcolor=blue,
    citecolor=blue,
}
\usepackage{comment}
\usepackage{pdflscape}
\usepackage{placeins}
\usepackage{scalefnt}
\usepackage{xcolor}                 % For colors
\usepackage{relsize}

\newcommand{\Msun}{\hbox{$\mathrm{M}_{\rm{\odot}}$}}

\newcommand{\cigale}{\hbox{\texttt{CIGALE}}\xspace}
\newcommand{\prospector}{\hbox{\texttt{Prospector}}\xspace}
\newcommand{\redback}{\hbox{\texttt{Redback}}\xspace}

\newcommand{\halpha}{\hbox{${\mathrm{H}}{\alpha}$}\xspace}
\newcommand{\hbeta}{\hbox{${\mathrm{H}}{\beta}$}\xspace}
\newcommand{\hgamma}{\hbox{${\mathrm{H}}{\gamma}$}\xspace}

\newcommand{\OIIab}{\hbox{[O\,\textsc{ii}]\,$\lambda\lambda3727,3729$}\xspace}

\newcommand{\OIIIb}{\hbox{[O\,\textsc{iii}]\,$\lambda5007$}\xspace}

\newcommand{\NIIa}{\hbox{[N\,\textsc{ii}]\,$\lambda6548$}\xspace}
\newcommand{\NIIb}{\hbox{[N\,\textsc{ii}]\,$\lambda6583$}\xspace}
\newcommand{\SIIa}{\hbox{[S\,\textsc{ii}]\,$\lambda6717$}\xspace}

\newcommand{\paeight}{\hbox{${\mathrm{Pa}}8$}\xspace}
\newcommand{\pagamma}{\hbox{${\mathrm{Pa}}{\gamma}$}\xspace}
\newcommand{\pabeta}{\hbox{${\mathrm{Pa}}{\beta}$}\xspace}
\newcommand{\paalpha}{\hbox{${\mathrm{Pa}}{\alpha}$}\xspace}

\newcommand{\brdelta}{\hbox{${\mathrm{Br}}{\delta}$}\xspace}
\newcommand{\brgamma}{\hbox{${\mathrm{Br}}{\gamma}$}\xspace}
\newcommand{\brbeta}{\hbox{${\mathrm{Br}}{\beta}$}\xspace}

\newcommand{\orcid}[1]{\href{https://orcid.org/#1}{\includegraphics[width=8pt]{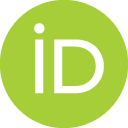}}}

\graphicspath{{figures/}} %Setting the graphicspath

\begin{document}
    \title{JWST unveils a dust-obscured supernova associated with a gamma-ray burst}

\author{
B.~Schneider\orcid{0000-0003-4876-7756}\inst{1}\fnmsep\thanks{E-mail: benjamin.schneider@lam.fr},
A.~J.~Levan\orcid{0000-0001-7821-9369}\inst{2,3},
N.~Sarin\orcid{0000-0003-2700-1030}\inst{4,5},
N.~A.~Rakotondrainibe\orcid{0009-0004-0263-7766}\inst{1},
B.~P.~Gompertz\orcid{0000-0002-5826-0548}\inst{6,7},
V.~Buat\orcid{0000-0003-3441-903X}\inst{1},
J.~T.~Palmerio\orcid{0000-0002-9408-1563}\inst{8},
D.~B.~Malesani\orcid{0000-0002-7517-326X}\inst{9,10,2},
A.~de~Ugarte~Postigo\orcid{0000-0001-7717-5085}\inst{1},
A.~Martin-Carrillo\orcid{0000-0001-5108-0627}\inst{11},
A.~Rossi\orcid{0000-0002-8860-6538}\inst{12}, \\
V.~D'Elia\orcid{0000-0003-3703-4418}\inst{13},
M.~De~Pasquale\orcid{0000-0002-4036-7419}\inst{14},
D.~Hartmann\orcid{0000-0002-8028-0991}\inst{15},
P.~Jakobsson\orcid{0000-0002-9404-5650}\inst{16},
E.~Le~Floc'h\orcid{0000-0001-7421-4413}\inst{8},
P.~O'Brien\orcid{0000-0002-5128-1899}\inst{17}, \\
E.~Pian\orcid{0000-0001-8646-4858}\inst{12},
G.~Pugliese\orcid{0000-0003-3457-9375}\inst{18},
A.~Raidani\orcid{0000-0001-5980-0174}\inst{19,20},
R.~Salvaterra\orcid{0000-0002-9393-8078}\inst{21},
J.~Sollerman\orcid{0000-0003-1546-6615}\inst{22},
N.~R.~Tanvir\orcid{0000-0003-3274-6336}\inst{17}, \\
S.~D.~Vergani\orcid{0000-0001-9398-4907}\inst{23},
\and D.~Watson\orcid{0000-0002-4465-8264}\inst{9,10}
}

\institute{
%1
Aix Marseille University, CNRS, CNES, LAM, Marseille, France
\and %2
Department of Astrophysics/IMAPP, Radboud University, PO Box 9010, 6500 GL Nijmegen, The Netherlands
\and %3
Department of Physics, University of Warwick, Gibbet Hill Road, CV4 7AL Coventry, United Kingdom
\and %4
Kavli Institute for Cosmology, University of Cambridge, Madingley Road, CB3 0HA, UK
\and %5
Institute of Astronomy, University of Cambridge, Madingley Road, CB3 0HA, UK
\and %6
School of Physics and Astronomy, University of Birmingham, Birmingham, B15 2TT, UK
\and %7
Institute for Gravitational Wave Astronomy, University of Birmingham, Birmingham, B15 2TT, UK
\and %8
Université Paris-Saclay, Université Paris Cité, CEA, CNRS, AIM, 91191, Gif-sur-Yvette, France
\and %9
Cosmic Dawn Center (DAWN), Denmark
\and %10
Niels Bohr Institute, University of Copenhagen, Jagtvej 155A, 2200 Copenhagen N, Denmark
\and %11
School of Physics and Centre for Space Research, University College Dublin, Belfield, Dublin 4, Ireland
\and %12
INAF -- Osservatorio di Astrofisica e Scienza dello Spazio, Via Piero Gobetti 93/3, 40129 Bologna, Italy
\and %13
ASI Space Data Centre, Via del Politecnico snc, 00133 Rome, Italy
\and %14
University of Messina, Mathematics, Informatics, Physics and Earth Science Department, Via F.S. D'Alcontres 31, 98166, Messina, Italy
\and %15
Clemson University, Department of Physics and Astronomy, Clemson, SC 29634-0978, USA
\and %16
Centre for Astrophysics and Cosmology, Science Institute, University of Iceland, Dunhagi 5, 107, Reykjavik, Iceland
\and %17
School of Physics and Astronomy, University of Leicester, University Road, Leicester, LE1 7RH, UK
\and %18
Anton Pannekoek Institute of Astronomy, University of Amsterdam, Science Park 904, 1098 XH Amsterdam, The Netherlands
\and %19
School of Mathematical and Physical Sciences, Macquarie University, NSW 2109, Australia
\and %20
Macquarie University Astrophysics and Space Technologies Research Centre, Sydney, NSW 2109, Australia
\and %21
INAF -- Istituto di Astrofisica Spaziale e Fisica Cosmica di Milano, Via A. Corti 12, 20133 Milano, Italy
\and %22
Oskar Klein Centre, Department of Astronomy, Stockholm University, AlbaNova, SE-106 91 Stockholm, Sweden
\and %23
LUX, Observatoire de Paris, Université PSL, CNRS, Sorbonne Université, 92190 Meudon, France
}

    % These dates will be filled out by the publisher
    \date{Accepted XXX. Received YYY; in original form ZZZ}

  \abstract
   {Recent observations have challenged the duration-based classification of gamma-ray bursts (GRBs), making the detection of an associated supernova (SN) or kilonova crucial for conclusively identifying their progenitors.}
   {We investigate the origin of GRB~240825A at $z=0.659$, whose prompt emission gave conflicting indications of a merger or collapsar origin and for which no SN was detected in deep ground-based observations.}
   {We analyze James Webb Space Telescope (JWST) NIRSpec observations of GRB~240825A obtained 66.5~days (40.1~days in the rest frame) after the burst. We identify a SN signal using a pixel-by-pixel decomposition of the 2D spectrum and isolate its emission through a joint modeling of the host and transient emission. We constrain the line of sight extinction from the near-infrared-to-X-ray afterglow spectral energy distribution and characterize the host through broadband photometry and nebular emission lines.}
   {Despite a dusty line of sight ($A^{\rm GRB}_V=1.37\pm0.08$~mag), with tentative evidence of a 2175~\AA\ extinction bump, and a bright host galaxy ($M_B = -20.5$~AB mag), we identify a point-like source at the GRB position with broad spectral features similar to those of the archetypal SN~1998bw.
   The SN contributes $\sim$10--15\% of the NIRSpec flux and would contribute only $\sim$1\% of the total flux in a ground-based optical image at the same epoch. 
   Our spectral decomposition indicates a SN with a luminosity corrected for extinction comparable to SN~1998bw, although the absolute luminosity remains sensitive to the host subtraction and template assumptions. The host is an extended, dusty, actively star-forming galaxy with $\log(M_\star/M_\odot)=10.12\pm0.05$, a star formation rate of $5.88\pm1.60\,{\rm M_\odot}\,\mathrm{yr}^{-1}$, and a super-solar gas-phase metallicity of $12+\log({\rm O/H})=8.87\pm0.04$, above the metallicity typically observed for collapsar GRB hosts.}
   {The detection of a dust-obscured SN firmly establishes the massive-star origin of GRB~240825A and shows that JWST can uncover and extend the study of SNe in dusty, chemically enriched environments that were previously largely inaccessible.}

   \titlerunning{JWST unveils a dust-obscured SN associated with a GRB}
   \authorrunning{B.~Schneider et al.}
   \maketitle
%

%-------------------------------------------------------------------
\section{Introduction}
\label{sec:Introduction} 
% \linenumbers

Gamma-ray bursts (GRBs) are brief, intense flashes of gamma rays and are among the most luminous transients in the Universe. The duration of their prompt emission ($T_{90}$) reveals at least two main populations, historically separated at $T_{90}\sim2$~s \citep[e.g.,][]{kouveliotou1993a}, with short GRBs typically showing harder spectra and long GRBs softer spectra.
This dichotomy has been widely interpreted as tracing two distinct progenitor channels. Long GRBs are thought to originate from the core collapse of massive, rapidly rotating stars, a scenario strongly supported by the spatial and temporal association of several GRBs with broad-lined Type Ic supernovae \citep[SNe Ic-BL; e.g.,][]{galama1998a,hjorth2003a,cano2017a}.
In contrast, short GRBs are attributed to mergers of compact-object binaries, such as neutron stars, a scenario supported in one case by the association with gravitational-wave events \citep{abbott2017a} and by the identification of kilonova (KN) emission \citep[e.g.,][]{tanvir2013a,levan2023b}.

However, this classification has been challenged by a growing number of events that lie in the transition region between the two populations or exhibit properties inconsistent with a purely duration-based criterion \citep{rastinejad2022a,troja2022a,yang2022a,gompertz2023a,levan2023b}. Establishing the progenitors of GRBs and building well-classified samples therefore remain central goals for understanding the physics of these explosions \citep[e.g.,][]{levan2026a}. Such samples are essential for assessing the contribution of compact-object mergers to heavy-element enrichment and for understanding how GRB progenitors and explosion mechanisms evolve across cosmic time.

The most robust way to confirm the progenitor of a GRB is through spectroscopic observations of the associated SN or KN. However, such observations are expensive and, from the ground, largely limited to relatively nearby events \citep[$z \lesssim 0.5$; e.g.,][]{finneran2025a}. A complementary approach is repeated optical and near-infrared (NIR) imaging to monitor the afterglow light curve and search for an emerging SN. This method has yielded photometric detections of SNe associated with GRBs up to $z \sim 1$ \citep[e.g.,][]{tanvir2010a,finneran2025a}, although it lacks the spectral information needed to fully characterize the SN. In both approaches, the SN signal, which typically peaks $\sim$10--15~days (rest-frame) after the burst \citep{cano2017a}, can be strongly diluted by a slowly decaying afterglow, a bright host galaxy, or both. This can make the presence of a SN ambiguous and leave the progenitor classification uncertain.

GRB~240825A, at $z=0.659$, was the subject of extensive multiwavelength follow-up owing to its bright prompt and afterglow emission. Analyses of its prompt emission reveal an additional hard spectral component beyond the usual sub-MeV emission, with differing interpretations of its spectral break and evidence of a thermal contribution \citep{zhang2025a,wang2025a,wu2025a}.
A quasi-periodic modulation of the photospheric emission has also been reported \citep{li2025d}, and neutrino non-detections have been used to constrain the jet parameters under different emission models \citep{pradhan2026a}. The optical and NIR observations have been interpreted as early reverse-shock emission followed by forward-shock emission \citep{wu2025a}. The unusually red afterglow indicates substantial dust extinction along the line of sight \citep{cheng2025a,li2026b}.
Despite this extensive follow-up, the progenitor of GRB~240825A remained ambiguous \citep{gupta2026a}. Its relatively short, hard prompt burst ($T_{90}\simeq4$~s in the 50--300~keV band), near the short--long boundary observed by the \textit{Fermi} Gamma-ray Burst Monitor \citep[GBM;][]{meegan2009a}, and rapid variability ($\sim14$~ms) suggested a possible merger origin. A soft extended tail following the main pulse was also detected by the Burst Alert Telescope \citep[BAT;][]{barthelmy2005a} on board the Neil Gehrels \textit{Swift} Observatory \citep{gehrels2004a}, with a duration of $\sim57$~s. In contrast, its high isotropic-equivalent energy ($E_{\gamma,\mathrm{iso}}\simeq1.6\times10^{53}$~erg) and position on the Amati relation \citep{amati2002a} favored a massive-star origin. Deep optical and NIR follow-up revealed no associated SN, with limits reaching $r>25.0$ and $z>24.8$~mag at 17.6~days after the burst, leaving an intrinsically faint or dust-obscured event among the possible scenarios \citep{gupta2026a}.

The sensitivity and infrared coverage of the James Webb Space Telescope (JWST; \citealt{gardner2006a}) make it a powerful facility for detecting and identifying SNe. 
The Near Infrared Spectrograph \citep[NIRSpec;][]{jakobsen2022a} on board JWST has enabled the spectroscopic identification of SNe Ic-BL associated with GRB~221009A at late times \citep{blanchard2024a} despite huge foreground extinction that initially challenged even JWST \citep{levan2023a}. 
It also identified a SN in the candidate orphan afterglow AT2023lcr/SN~2023lcr at $z=1.03$ \citep{martin-carrillo2023a,li2025a}, and XRF~241001A/SN~2024aiiq at $z=0.573$ \citep{schneider2026b}, pushing spectroscopic studies of GRB/SNe to higher redshifts than are accessible from the ground ($z \lesssim 0.5$).
Such studies now extend to extragalactic fast X-ray transients discovered by Einstein Probe \citep[EP;][]{yuan2022a}, including infrared spectroscopy of EP250108a/SN~2025kg \citep{rastinejad2025a} and the photometric and spectroscopic identification of EP240801a/SN~2024aihh at $z=1.67$ \citep{vanhoof2026a}. Imaging with the JWST Near Infrared Camera \citep[NIRCam;][]{rieke2023a} has also uncovered numerous high-redshift SN candidates in deep extragalactic surveys \citep{decoursey2025a} and detected the SN associated with GRB~250314A at $z\simeq7.3$ \citep{levan2025a}. Together, these observations extend studies of the connection between high-energy transients and stellar explosions across cosmic time and enable the identification of SNe that are obscured by dust or diluted by afterglow and host emission.

In this work, we present JWST/NIRSpec observations of GRB~240825A obtained 66.5~days (40.1~days in the rest frame) after the trigger, and identify an associated, dust-obscured SN Ic-BL. By separating the transient emission from the underlying host emission, we establish the massive-star origin of the burst, despite the absence of a SN detection in earlier ground-based observations. We also constrain the line of sight extinction and characterize the host galaxy.
Section~\ref{sec:Observation} describes the observations and data reduction. Section~\ref{sec:Data_analysis} presents the afterglow modeling, the SN identification and spectral decomposition, and the host galaxy analysis. In Sect.~\ref{sec:Discussion_conclusions}, we discuss the implications for GRB classification and summarize our conclusions.

Throughout the paper, we adopt a flat $\Lambda$ cold dark matter ($\Lambda$CDM) cosmology from \citet{planckcollaboration2020a}, with $\Omega_{\rm m} = 0.315$, $\Omega_{\Lambda} = 0.685$, and $H_{0} = 67.4$~km~s$^{-1}$~Mpc$^{-1}$. All magnitudes are reported in the AB system, and uncertainties are quoted at the $1\sigma$ level unless otherwise stated. 
 
%--------------------------------------------------------------------
\section{Observations and data reduction}
\label{sec:Observation}

\subsection{Optical and near-infrared photometry}
The afterglow of GRB~240825A was followed extensively from the ground and from space \citep[e.g.,][]{wu2025a,gupta2026a}. In this work, we use the early observations obtained with the \textit{Swift} Ultraviolet/Optical Telescope \citep[UVOT;][]{roming2005a} reported by \citet{gupta2026a}. We also use the $g'$, $r'$, and $z'$ photometry obtained just before the X-shooter spectrum \citep{wu2025a} with the acquisition camera of the X-shooter spectrograph \citep{vernet2011a}, mounted on the Very Large Telescope (VLT) of the European Southern Observatory (ESO).

At the afterglow position, archival images reveal an extended source, including those of the DESI Legacy Surveys DR10 \citep{dey2019a} and the Hyper Suprime-Cam Subaru Strategic Program \citep[HSC-SSP;][]{aihara2022a}. 
This spatial coincidence, together with the strong emission lines detected in the X-shooter spectrum \citep{wu2025a} at the same redshift as the afterglow absorption features, supports its identification as the host galaxy.
Photometric observations of the host are available from archival surveys and the literature.
At $\gtrsim$25~days after the trigger, \citet{wu2025a} detected the host with HiPERCAM \citep{dhillon2021a} mounted on the 10.4-m Gran Telescopio Canarias (GTC) and \citet{gupta2026a} reported NIR detections with the Espectrógrafo Multiobjeto Infra-Rojo \citep[EMIR;][]{garzon2022a}, also mounted on the GTC.

In our analysis, we favor host photometry obtained with a single instrument across several bands.
We use the HiPERCAM $u_s,g_s,r_s,i_s,z_s$ observations acquired 33.4~days after the burst \citep{wu2025a} and the GTC/EMIR $J,H,K$ data obtained at 27.4~days \citep{gupta2026a}.
We supplement these with archival pre-burst observations, a HSC-SSP $y$-band image retrieved from the HSC Legacy Archive and WISE observations in the $W1$, $W2$, and $W3$ bands \citep{wright2010a}.
We perform aperture photometry on the reduced images using {\sc stdweb} \citep{karpov2025a}. 
The $u_s$ band is calibrated against the Sloan Digital Sky Survey (SDSS) catalog \citep{york2000a,gunn2006a,padmanabhan2008a}, the other optical bands against the Pan-STARRS1 (PS1) catalog \citep{chambers2016a,magnier2020a,flewelling2020a}, and the NIR bands against the Two Micron All Sky Survey (2MASS) catalog \citep{skrutskie2006a}. For WISE, we adopt the measurements from the DESI Legacy Surveys DR10 catalog \citep{dey2019a}.
The resulting magnitudes are listed in Table~\ref{tab:grb240825a_host_photometry}.

\subsection{Optical and near-infrared spectroscopy}

Spectroscopic observations of the optical counterpart were obtained with the VLT/X-shooter instrument on 2024-08-26, 11.35~hr after the trigger, and were presented by \citet{wu2025a}. In addition to the afterglow continuum, the spectrum shows strong emission lines (Fig.~\ref{fig:xshooter_spec}) that we use to characterize the host galaxy. We also use the spectrum to construct the NIR-to-X-ray spectral energy distribution (SED) and measure the extinction along the GRB line of sight.

The absolute flux calibration of the X-shooter spectrum can be affected by several sources of uncertainty, including slit losses, which are expected to be significant given the seeing of $\approx$1.2$''$ measured in the $r'$ band acquisition image.
We therefore rescale the spectrum to the $g'$, $r'$, and $z'$ magnitudes reported by \citet{wu2025a} obtained prior to the spectrum. The scaling factor is determined by minimizing the differences between the observed magnitudes and the synthetic magnitudes computed from the spectrum. This yields a factor of 2.15, which we apply to all three arms throughout the subsequent analysis.

\begin{figure}[t]
    \centering
    \includegraphics[width=\hsize]{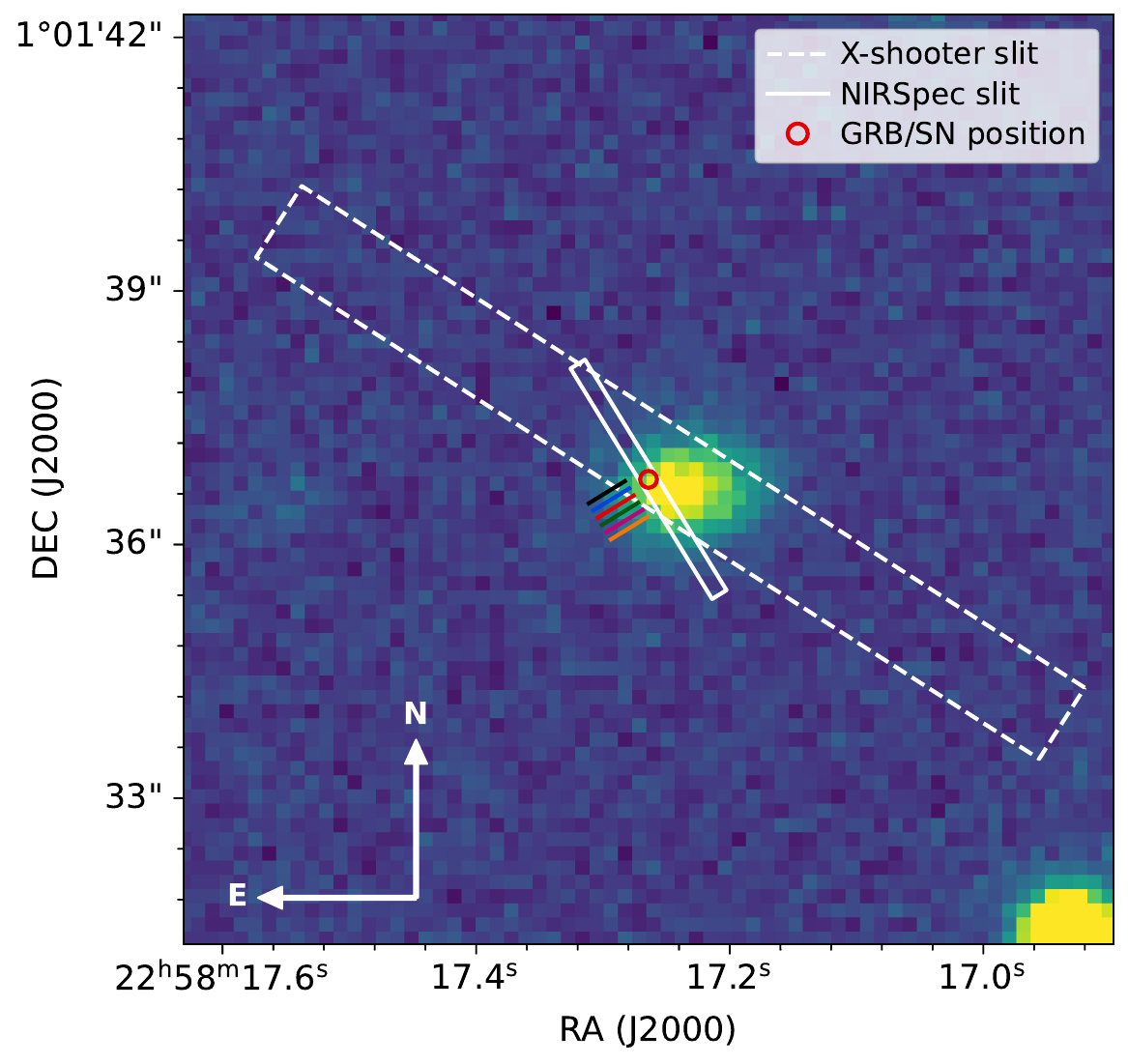} % \hsize
    \caption{Subaru/HSC $r$ band image of the GRB~240825A host galaxy. The white solid and dashed rectangles indicate the JWST/NIRSpec and VLT/X-shooter slit positions, respectively. The red circle marks the best GRB/SN position. Colored lines along the NIRSpec slit mark the extraction positions of the 1D spectra shown in Fig.~\ref{fig:jwst_spectrum_line_by_line}. North is up, and east is to the left.}
    \label{fig:slitpos}
\end{figure}

About two months later, we observed the position of GRB~240825A with the JWST/NIRSpec instrument under program GO~6133 (PI: Gompertz). The observations were performed on 2024-10-31 at 03:24:38~UT, 66.5~days (40.1~days in the rest frame) after the \textit{Fermi} trigger, using the PRISM/CLEAR mode and the S200A1 fixed slit ($0.2\arcsec\times3.2\arcsec$), for an effective exposure time of 1.7~hr.
The slit was centered on the best-known afterglow position (RA = $344.57193^\circ$, Dec = $1.02688^\circ$; Fig.~\ref{fig:slitpos}) using a blind offset.
We use the standard level~3 pipeline products \citep{bushouse2023a} retrieved from the Mikulski Archive for Space Telescopes (MAST).

%--------------------------------------------------------------------
\section{Data analysis and results}
\label{sec:Data_analysis}

\subsection{Afterglow SED modeling}
\label{ssec:afterglow_sed}
We determine the shape of the NIR-to-X-ray afterglow continuum by constructing an SED at a mid-time of $\sim$11.2~hr after the trigger, matching the quasi-simultaneous VLT/X-shooter and \textit{Swift} X-Ray Telescope \citep[XRT;][]{burrows2005a} observations. We further extend the SED into the ultraviolet (UV) using the \textit{Swift}/UVOT observations reported by \citet{gupta2026a}.

\begin{figure}[t]
    \centering
    \includegraphics[width=\hsize]{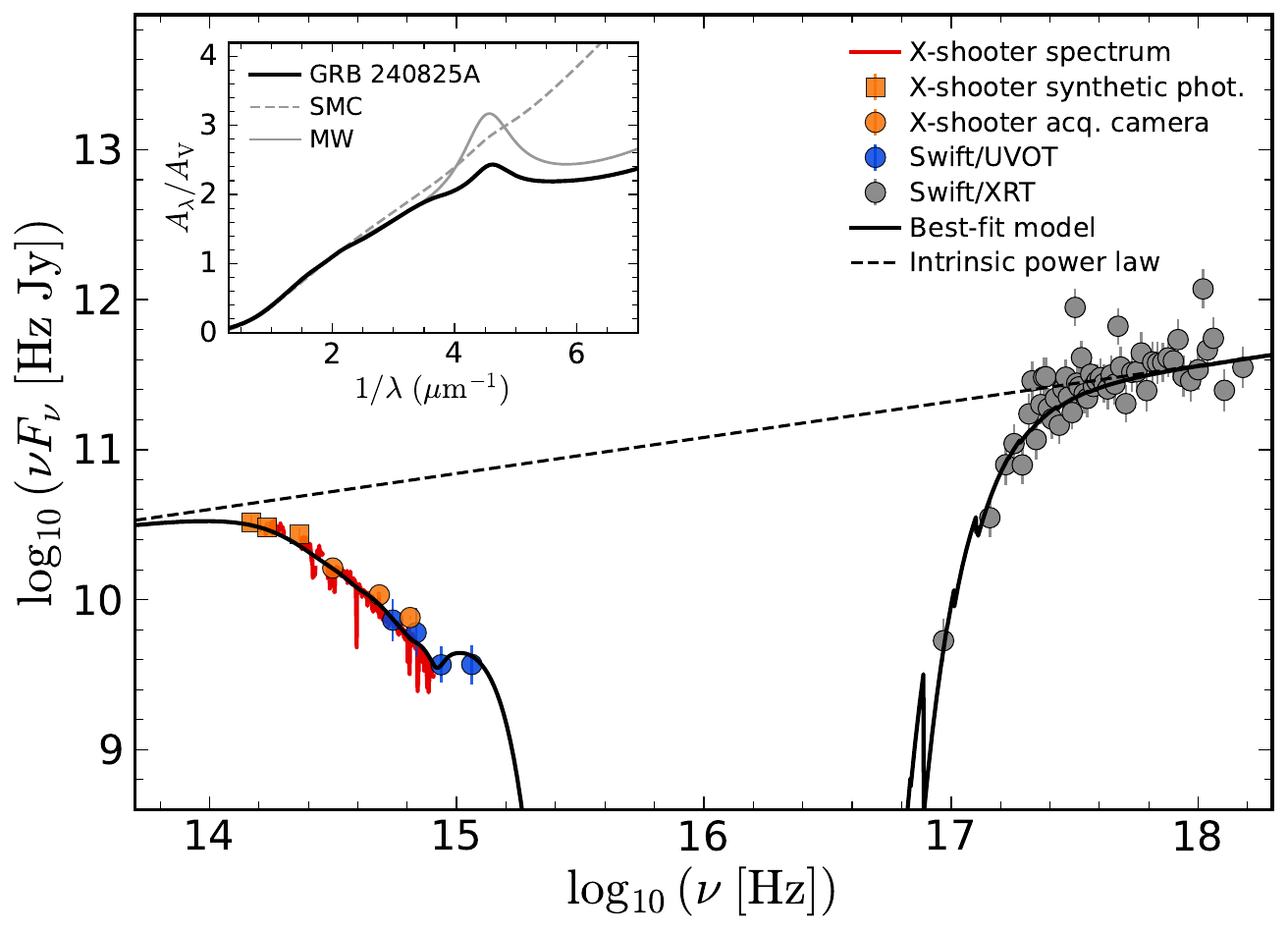} % \hsize
    \caption{NIR-to-X-ray SED of the GRB~240825A afterglow $\sim$11.2~hr after the trigger. The red line shows the X-shooter spectrum and orange squares synthetic photometry derived from it. Blue circles show the \textit{Swift}/UVOT photometry extrapolated to the SED epoch using $\alpha = -0.94 \pm 0.01$ \citep{gupta2026a}, and gray circles the \textit{Swift}/XRT spectrum. The solid black curve shows the best-fit model obtained using the spectroscopic data set, and the dashed line the intrinsic power law. The inset compares the extinction curve measured for GRB~240825A (black) with the SMC (gray dashed) and MW (gray solid) laws.}
    \label{fig:photospectroSED}
\end{figure}

For the X-ray data, we extract the 0.3--10~keV XRT spectrum over $T_0 + 6.11$--$16.39$~h, centered on the SED epoch and overlapping with the X-shooter observations, using the automated \texttt{time-sliced spectra} tool, and bin it to a minimum of 10 counts per bin. 
For the optical-to-NIR data, we combine the X-shooter UVB, VIS, and NIR spectra, excluding the low signal-to-noise ratio (S/N) edges and wavelengths affected by telluric absorption.
No UVOT observations are available at 11.2~h, but \citet{gupta2026a} reported earlier detections in the $U$, $B$, $V$, and $UVW1$ filters. We extrapolate these measurements to the SED epoch using the post-break optical decay index of $\alpha = -0.94 \pm 0.01$ derived by \citet{gupta2026a}. Since the $B$ and $V$ bands overlap with the X-shooter wavelength coverage, we verify the extrapolation by comparing the extrapolated fluxes with the X-shooter spectrum and find good agreement (Fig.~\ref{fig:photospectroSED}).
All data are corrected for Galactic extinction ($A_{\rm V}^{\rm Gal} = 0.17$~mag; \citealt{schlafly2011a}) and Galactic X-ray absorption ($N_{\rm H}^{\rm Gal} = 5.28 \times 10^{20}$~cm$^{-2}$; \citealt{bekhti2016a}).

\begin{figure*}[t]
    \sidecaption
    \includegraphics[width=12cm]{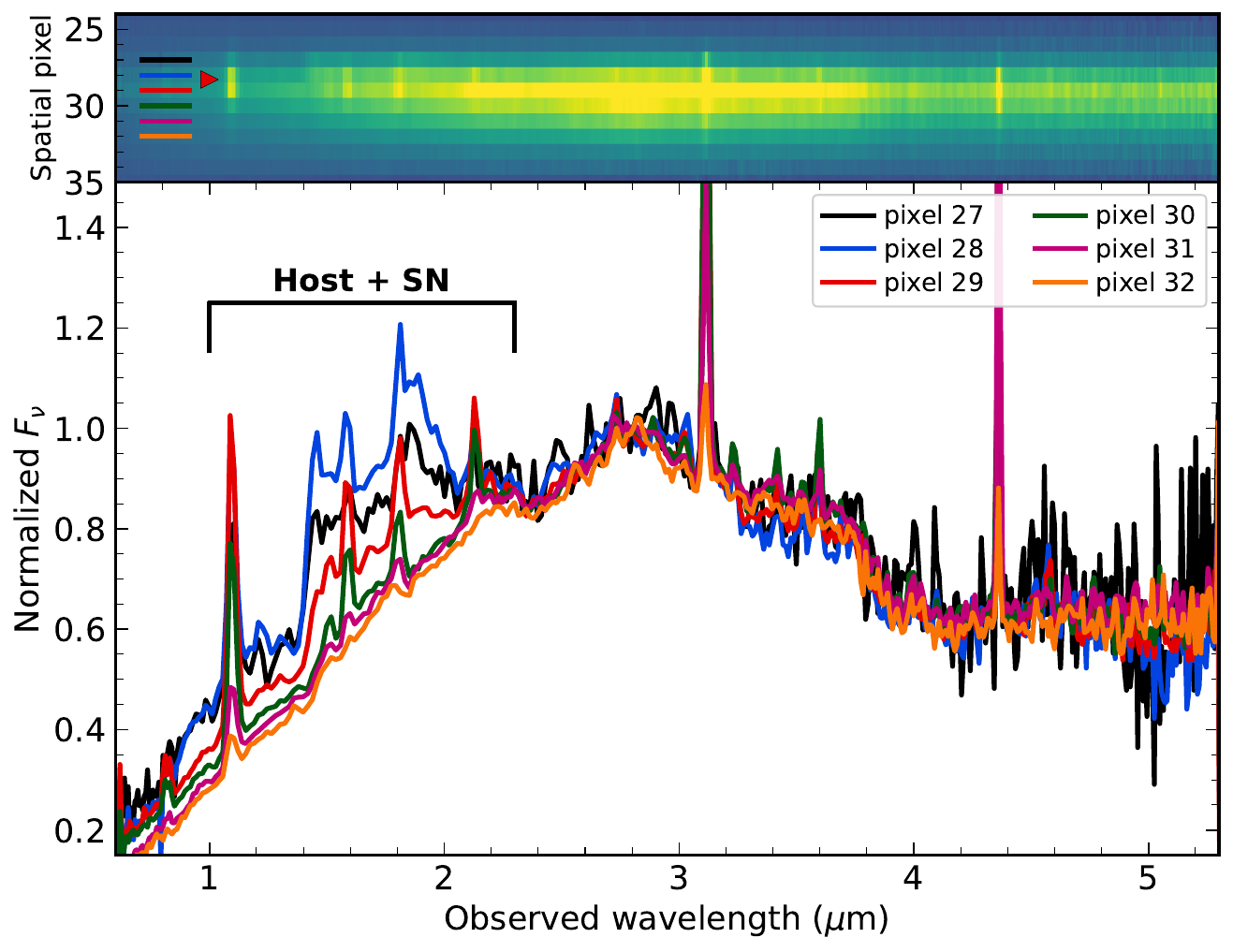} % \hsize
    \caption{JWST/NIRSpec observations of GRB~240825A 66.5~days (40.1~days in the rest frame) after the trigger. Top: 2D spectrum, with the expected GRB/SN position marked by a red triangle (Pixel 28.28). Colored horizontal markers indicate the spatial pixels (27--32) from which the 1D spectra shown in the bottom panel were extracted. The black line marks the region where the spectra show combined host and SN emission. The spectra are normalized at $\lambda_{\rm obs}=2.9\,\mu{\rm m}$ for visibility.}
    \label{fig:jwst_spectrum_line_by_line}
\end{figure*}

From these data, we build two SED data sets. The first is spectroscopic and consists of the X-shooter spectrum in the optical-NIR. The second is purely photometric and combines the \textit{Swift}/UVOT photometry, extrapolated to the SED epoch, with the X-shooter acquisition camera photometry ($g'$, $r'$, $z'$) and synthetic photometry derived from the X-shooter spectrum.
We fit both data sets independently, together with the XRT spectrum, using a Markov chain Monte Carlo (MCMC) algorithm to estimate the intrinsic spectral parameters of the afterglow. For each data set, we test both a single and a broken power-law model for the continuum, combined with the flexible dust extinction parameterization of \citet{fitzpatrick1990a} (hereafter FM).
The FM law consists of (1) a linear UV component described by the intercept $c_1$ and slope $c_2$; (2) a Lorentzian-like Drude profile accounting for the 2175~\AA\ bump, where $c_3$ sets the height, $\gamma$ the width, and $x_0$ the central position; and (3) a far-UV curvature term defined by $c_4$. To limit degeneracies, better constrain the remaining parameters, and derive the bump strength, we fix the width and central position of the Drude profile to $\gamma = 1\,\mu\mathrm{m}^{-1}$ and $x_0 = 4.6\,\mu\mathrm{m}^{-1}$, respectively \citep{fitzpatrick1999a}. 
For comparison, we also perform fits using the standard Milky Way (MW), Large Magellanic Cloud (LMC), and Small Magellanic Cloud (SMC) extinction laws \citep{pei1992a}.

Both models provide a good fit to the data. To compare them, we use the Bayesian information criterion (BIC; \citealt{schwarz1978a}), which penalizes models with additional free parameters. The broken power law is not statistically preferred ($\Delta{\rm BIC} \approx 0.6$) and we adopt the single power law model in the rest of our analysis.
The best-fit model using the single power-law is obtained with the FM extinction curve ($\chi^2_\mathrm{red}$ = 0.77) and is shown in Fig.~\ref{fig:photospectroSED}. It yields an intrinsic spectral slope of $\beta_{\rm OX} = 0.76_{-0.02}^{+0.01}$\footnote{The subscript follows the usual optical-to-X-ray notation, although the fit also includes the NIR data.} and a host galaxy visual extinction of $A^{\rm GRB}_V = 1.37_{-0.08}^{+0.08}$~mag. The best-fit models, including those obtained with the MW, LMC, and SMC laws, are listed in Table~\ref{tab:240825Afitsed}. Our measurements are consistent with previously reported values \citep{cheng2025a,li2026b,wu2025a,gupta2026a}, confirming substantial dust extinction along the line of sight.

Our models favor the presence of a 2175~\AA\ bump, as previously reported by \citet{li2026b}. This feature is suggested with both the spectroscopic and photometric data sets, although the two are not fully independent, since the latter includes synthetic photometry derived from the X-shooter spectrum.
To quantify the 2175~\AA\ bump, we compute its area, $A_{\mathrm{bump}} = \pi c_3/(2\gamma)$, and obtain $A_{\mathrm{bump}} = 1.59^{+1.05}_{-0.92}~\mu {\rm m}^{-1}$ ($1\sigma$), with $2\sigma$ and $3\sigma$ intervals of $^{+2.06}_{-1.25}$ and $^{+2.65}_{-1.50}$, respectively. 
Although the $1\sigma$ and $2\sigma$ intervals point to a detection, the lower bound of the $3\sigma$ interval approaches zero, which prevents a robust confirmation of the feature.
For the derived $A_V$, the measured bump strength is weaker than typically observed along MW \citep{fitzpatrick2007a} and Magellanic Cloud \citep{gordon2003a} sightlines, but it remains consistent with the overall trend of GRB afterglows showing this feature \citep{zafar2011a,zafar2012a,schady2012a,bolmer2018a,zafar2018c}.
While this feature is of interest for the dust properties along the line of sight, a more detailed analysis is beyond the scope of this paper and left for future work, as it does not affect the extinction correction or the results presented in the rest of our analysis.

\subsection{Supernova discovery}
The 2D NIRSpec spectrum (top panel of Fig.~\ref{fig:jwst_spectrum_line_by_line}) reveals an extended trace spanning several pixels, with strong emission lines characteristic of a star-forming galaxy. The 1D spectrum extracted from the full trace (Fig.~\ref{fig:jwst_spectrum_1D_lines}) is dominated by the host, with a continuum consistent with attenuated stellar emission and a rich set of nebular emission lines. We identify \hbeta, \OIIIb, \halpha, \paeight, \pagamma, \pabeta, \paalpha, \brdelta, \brgamma, and \brbeta, all at a consistent redshift of $z = 0.659$.

Taking advantage of the exquisite spatial resolution of JWST, we investigate how the spectral properties vary across the host galaxy. We use the \texttt{extract1d} step of the JWST pipeline \citep{bushouse2023a} to extract a 1D spectrum at each spatial pixel along the trace.
These pixel-by-pixel spectra, normalized at $\lambda_{\rm obs}=2.9\,\mu{\rm m}$, are shown in the bottom panel of Fig.~\ref{fig:jwst_spectrum_line_by_line}. Their extraction positions are indicated with the same color coding on the 2D spectrum (top panel of Fig.~\ref{fig:jwst_spectrum_line_by_line}) and on the Subaru/HSC image (Fig.~\ref{fig:slitpos}).
\label{ssec:empirical_host_sub}

\begin{figure*}[th!]
    % \sidecaption
    \centering
    \includegraphics[width=17cm]{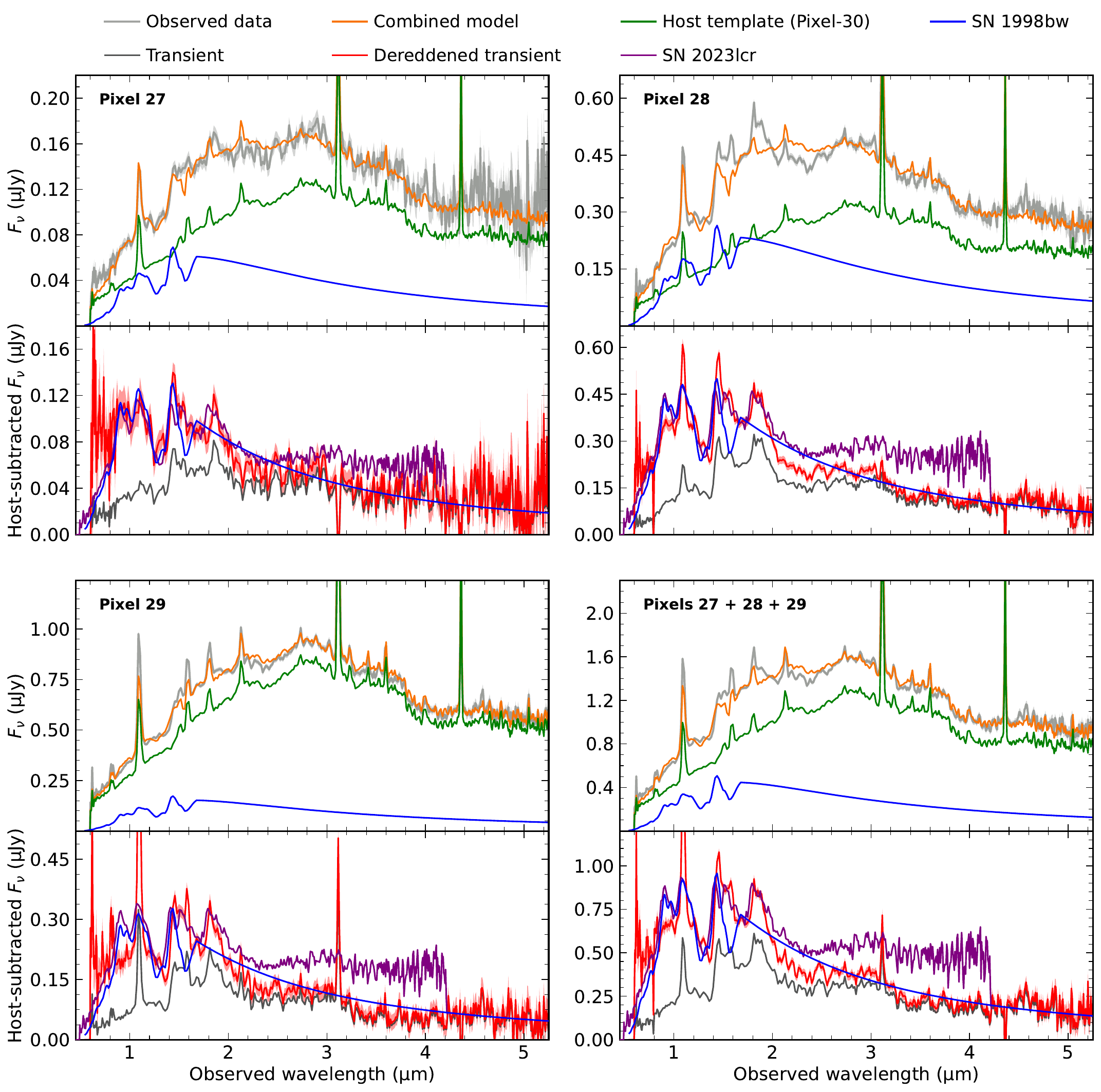} % \hsize
    \caption{Empirical modeling and host subtraction of the JWST/NIRSpec spectra of GRB~240825A for pixels 27 (upper left), 28 (upper right), and 29 (lower left), and their sum (lower right). In each panel, the upper subpanel shows the observed spectrum (gray) and the model (orange), which combines the scaled pixel 30 host spectrum (green) and the extrapolated SN~1998bw template attenuated by $A^{\rm GRB}_V$ (blue). The lower subpanel shows the host-subtracted spectrum before (dark gray) and after (red) correction for dust extinction ($A^{\rm GRB}_V$), together with the unattenuated SN~1998bw template (blue). The SN~2023lcr spectrum (purple), shifted to $z=0.659$, is independently scaled to each dereddened spectrum. Shaded regions indicate the $1\sigma$ uncertainties.}
    \label{fig:jwst_spectrum_joint_fit_pix30_sub}
\end{figure*}

The normalized 1D spectra have similar shapes at red wavelengths across the host galaxy. At bluer wavelengths ($\lesssim$2~$\mu$m), however, an additional component emerges, peaking at pixel 28 and spatially coincident with the GRB afterglow position (red triangle in the top panel of Fig.~\ref{fig:jwst_spectrum_line_by_line}). A fainter contribution is also detected in at least the two adjacent pixels (27 and 29), but no significant signal is observed beyond these pixels.
Given the spatial sampling of $0.1''$ ($\approx$0.72~kpc at $z=0.659$) per pixel and the JWST point spread function (PSF) full width at half maximum (FWHM) of $\sim$$0.1''$, an excess detected predominantly in a single pixel is consistent with an unresolved source.

While a more thorough analysis of the spectra is presented below, the spectral shape of the excess light presents broad features and is not consistent with the typical smooth, (extinguished) power-law behavior shown by GRB afterglows. Moreover, at the epoch of the JWST observations (66.5~days after the GRB), the afterglow is expected to have faded below the level required to account for the observed excess \citep[e.g.,][]{wu2025a}. This suggests that the excess arises from a SN associated with GRB~240825A, whose light curve would have peaked at $\sim$15~days in the rest frame and then declined progressively \citep[e.g.,][]{cano2017a}.

Interestingly, no SN signature was detected in earlier ground-based follow-up, despite deep optical and NIR observations \citep{gupta2026a}. The SN signal is also not clearly identifiable in the spatially integrated 1D spectrum (Fig.~\ref{fig:jwst_spectrum_1D_lines}), where it is diluted by the host emission. Its identification is made possible by the fine spatial resolution of JWST, which allows us to isolate the exact region where the SN-to-host contrast is highest.

\subsection{Empirical host galaxy subtraction}
To isolate the SN emission, we first use the spectrum of an adjacent spatial pixel as an empirical template for the local host emission, assuming that its SN contribution is negligible.
As shown in Fig.~\ref{fig:jwst_spectrum_line_by_line}, the pixel-30 spectrum contains no significant SN signal compared with pixels 27, 28, and 29, while exhibiting a consistent infrared continuum and similarly strong emission lines. 

To test whether pixel 30 probes a similar dust environment, we derive its extinction from the \halpha/\paalpha ratio, which is expected to be less affected by underlying stellar absorption than ratios involving higher-order Balmer lines. Assuming Case~B recombination with an electron temperature of $T_{\rm e}=10^4$~K, an electron density of $n_{\rm e}=10^2$~cm$^{-3}$, and the MW extinction law of \citet{pei1992a}, we obtain a host extinction of $A_V (\halpha/\paalpha) = 1.40 \pm 0.07$~mag for pixel 30, consistent with the extinction measured towards the GRB/SN (Sect.~\ref{ssec:afterglow_sed}).
Although the two measurements probe different dust geometries, with the nebular lines tracing \ion{H}{ii} regions and the GRB a single light-of-sight from the explosion site, this agreement suggests similar dust properties in the two regions (separated by $\sim$1~kpc), supporting the use of pixel 30 as a host template at the GRB/SN position.

We then fit the spectra of pixels 27, 28, and 29 simultaneously, each as the sum of a scaled pixel-30 spectrum and an extinguished SN~1998bw template. We adopt a common SN phase for the three pixels and allow the host scaling and SN normalization to vary independently for each pixel. We use the extrapolated SN~1998bw spectral templates implemented in \redback \citep{sarin2024a}, which are based on the spectral time series of \citet{patat2001a,vincenzi2019a,levan2026a} and extended in wavelength with a blackbody continuum. The template grid is shifted to the GRB redshift ($z=0.659$) and attenuated using the FM extinction law with $A^{\rm GRB}_{V}=1.37$~mag.

The upper subpanels of Fig.~\ref{fig:jwst_spectrum_joint_fit_pix30_sub} show the best-fit models and their individual components for the three spectra and their sum. The best fit is obtained for a SN phase of 54.9~days (33.1~days in the rest frame), with pixel-30 (host template) scaling factors of 0.11, 0.27, and 0.71 for pixels 27, 28, and 29, respectively. The corresponding SN amplitudes are 0.12, 0.45, and 0.30, giving a total amplitude of $0.85\pm0.01$ relative to SN~1998bw.

The joint fit reproduces the overall continuum well, although some spectral features, particularly in pixel 28, are not fully captured. SN~1998bw is not necessarily a perfect spectral analog of the SN associated with GRB~240825A, and, because of its low redshift, its observed spectra cover only part of the rest-frame wavelength range probed by NIRSpec at $z=0.659$. Outside this range, the templates rely on a blackbody extrapolation, which introduces additional uncertainty in the modeled continuum. The host emission lines are also not perfectly reproduced by the scaled pixel-30 spectrum. This may reflect spatial variations in nebular emission across the galaxy, as the different pixels probe physically distinct regions that may lie within or outside \ion{H}{ii} regions.

\begin{figure*}[t]
    \centering
    \includegraphics[width=17cm]{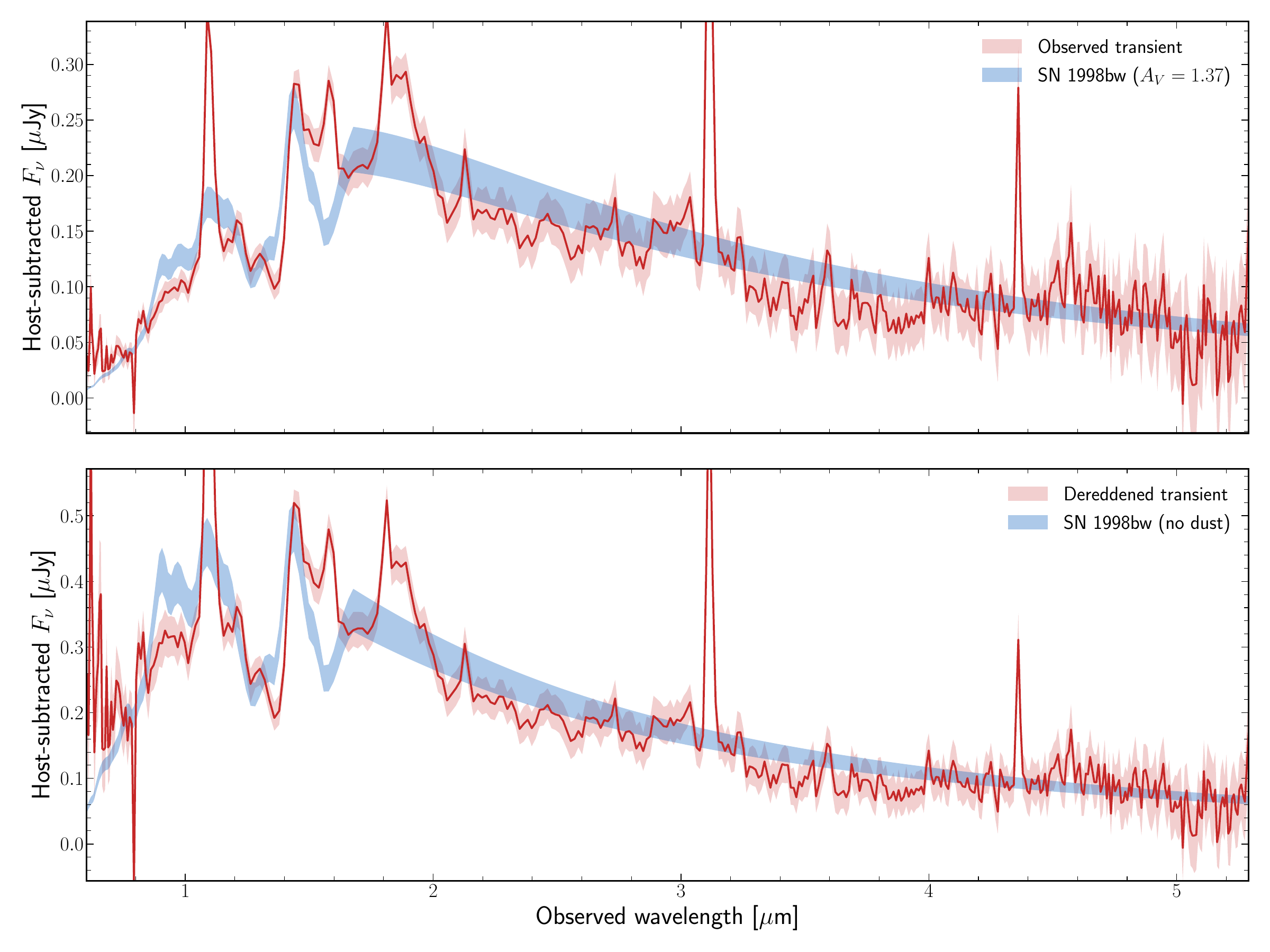}
    \caption{Posterior host--transient decomposition for pixel 28. The upper panel compares the host-subtracted signal with the extinguished SN~1998bw-like component. The lower panel shows the same quantities after dereddening the transient for $A^{\rm GRB}_{V}=1.37$~mag. Shaded regions contain 90\% of the posterior predictions. The median model curves are not shown. The narrow features are mostly nebular emission lines, which are not included in the smooth \prospector host model. The decomposition is derived from the joint fit to all spatial pixels, not from a pre-subtracted spectrum.}
    \label{fig:joint_pixel28}
\end{figure*}

The lower subpanels of Fig.~\ref{fig:jwst_spectrum_joint_fit_pix30_sub} show the host-subtracted spectra obtained after subtracting the scaled pixel-30 spectrum from pixels 27, 28, and 29, and from their sum. Assuming that the host-subtracted signal is dominated by the SN, we correct the spectra for dust extinction using the FM extinction law with $A_V^{\rm GRB}=1.37$~mag.
For comparison, we overplot the extrapolated SN~1998bw spectrum, shifted to the redshift of GRB~240825A and scaled by the amplitude derived from the joint fit above. We also compare the transient emission with the JWST/NIRSpec spectrum of SN~2023lcr \citep{martin-carrillo2023a}, a SN Ic-BL at $z=1.03$ associated with a candidate orphan GRB afterglow \citep{li2025a}. The SN~2023lcr spectrum is shifted to the redshift of GRB~240825A, accounting for cosmological effects, and normalized by fitting it to the host-subtracted spectra. We obtain scaling factors of 0.06, 0.26, and 0.19 for pixels 27, 28, and 29, respectively, and 0.51 for their sum.

The SN~2023lcr spectrum closely matches the shape of the transient emission, reproducing similar broad features and extending the agreement to redder wavelengths than the extrapolated SN~1998bw template. However, the SN~2023lcr spectrum shows an enhanced continuum at the reddest wavelengths that is not expected from a simple blackbody extrapolation. This may indicate a faint contribution from its host galaxy that emerges at redder wavelengths. Overall, the close spectral resemblance of the host-subtracted emission to both SN~1998bw and SN~2023lcr provides strong evidence that the SN associated with GRB~240825A is a Type Ic-BL.

While the comparison of spectral shapes is informative, the absolute normalizations should be interpreted with caution, because the two SNe were observed at different rest-frame phases. SN~2023lcr was observed $\sim$27~days after the likely explosion, while GRB~240825A was observed $\sim$40~days after the burst. Given the luminosity evolution of SNe Ic-BL over these timescales, the fitted normalization cannot be directly interpreted as the ratio of the intrinsic luminosities of the two events. Moreover, any SN contribution in pixel 30 would be partially removed by the host subtraction, leading to an underestimate of the recovered SN flux.

\subsection{Joint host--transient spectral decomposition}
\label{sec:joint_host_transient}
As a complement to the empirical subtraction based on a neighboring spatial pixel, we also model the galaxy and transient contributions simultaneously across all pixel spectra.
Unlike the empirical subtraction, this approach does not rely on a neighboring pixel as a host template, which may still contain transient light and affect the subtraction. It also propagates the uncertainty in the host continuum into the inferred transient spectrum.

We use the spectra extracted from pixels $i=26,\ldots,35$, retaining all finite flux measurements with positive uncertainties.
For each pixel spectrum, the observed flux density is modeled as,
\begin{equation}
\begin{aligned}
F_{\nu,i}(\lambda) ={}&
 b_i F_{\nu,{\rm gal}}(\lambda;\boldsymbol{\theta}_{\rm gal}) \\
&+ P_i(\mu_{\rm SN},\sigma_{\rm SN}) A_{{\rm SN},28}
 F_{\nu,{\rm 98bw}}(\lambda,t_{\rm SN})
 10^{-0.4 A_\lambda} \\
&+ \epsilon_i(\lambda),
\end{aligned}
\label{eq:joint_spectral_model}
\end{equation}
where $F_{\nu,\mathrm{gal}}(\lambda;\boldsymbol{\theta}_{\mathrm{gal}})$ is the galaxy spectrum, which is shared by all pixels and scaled by a pixel-dependent normalization $b_i$.
The galaxy spectrum is generated with the parametric star formation history model in \prospector/\texttt{FSPS} \citep{conroy2009a,conroy2010a,johnson2021a}.
Its free parameters ($\boldsymbol{\theta}_{\rm gal}$) are the stellar mass normalization, stellar metallicity, dust optical depth, stellar population age, and star formation timescale. The ten host normalizations are allowed to vary independently, with a Gaussian smoothness prior imposed on their second differences in log flux. This gives the galaxy sufficient spatial freedom without requiring its spectral shape to change between adjacent pixels.

The SN contribution is described by the SN~1998bw template at phase $t_{\mathrm{SN}}$, with normalization $A_{\mathrm{SN},28}$ and spatial profile $P_i(\mu_{\mathrm{SN}},\sigma_{\mathrm{SN}})$. The factor $10^{-0.4A_\lambda}$ accounts for extinction, and $\epsilon_i(\lambda)$ represents the residual noise.
The unresolved transient is assigned a wavelength-independent Gaussian spatial profile,
\begin{equation}
 P_i(\mu_{\rm SN},\sigma_{\rm SN}) \propto
 \exp\left[-\frac{(i-\mu_{\rm SN})^2}
 {2\sigma_{\rm SN}^2}\right],
 \label{eq:transient_spatial_profile}
\end{equation}
whose centroid and width are inferred from the data.
For simplicity, we assume this profile to be independent of wavelength, although the NIRSpec PSF broadens towards longer wavelengths.
We normalize the spatial profile to pixel 28, such that $P_{28}=1$, and quote the corresponding template normalization $A_{{\rm SN},28}$ for that pixel.
This profile is a phenomenological description of the compact component, not an imposed NIRSpec PSF.
As in Sect.~\ref{ssec:empirical_host_sub}, we use the SN~1998bw spectral time series implemented in \redback \citep{sarin2024a}, including the blackbody extrapolation beyond the wavelength coverage of the original template~\citep{patat2001a,vincenzi2019a,levan2026a}.
The template is consistently transformed to the redshift of the source ($z=0.659$) in observer-frame time, wavelength, and flux density. The normalization quoted is defined relative to SN~1998bw before applying the host extinction to the GRB.

\begin{figure*}[ht]
    \sidecaption
    % \centering
    \includegraphics[width=12cm]{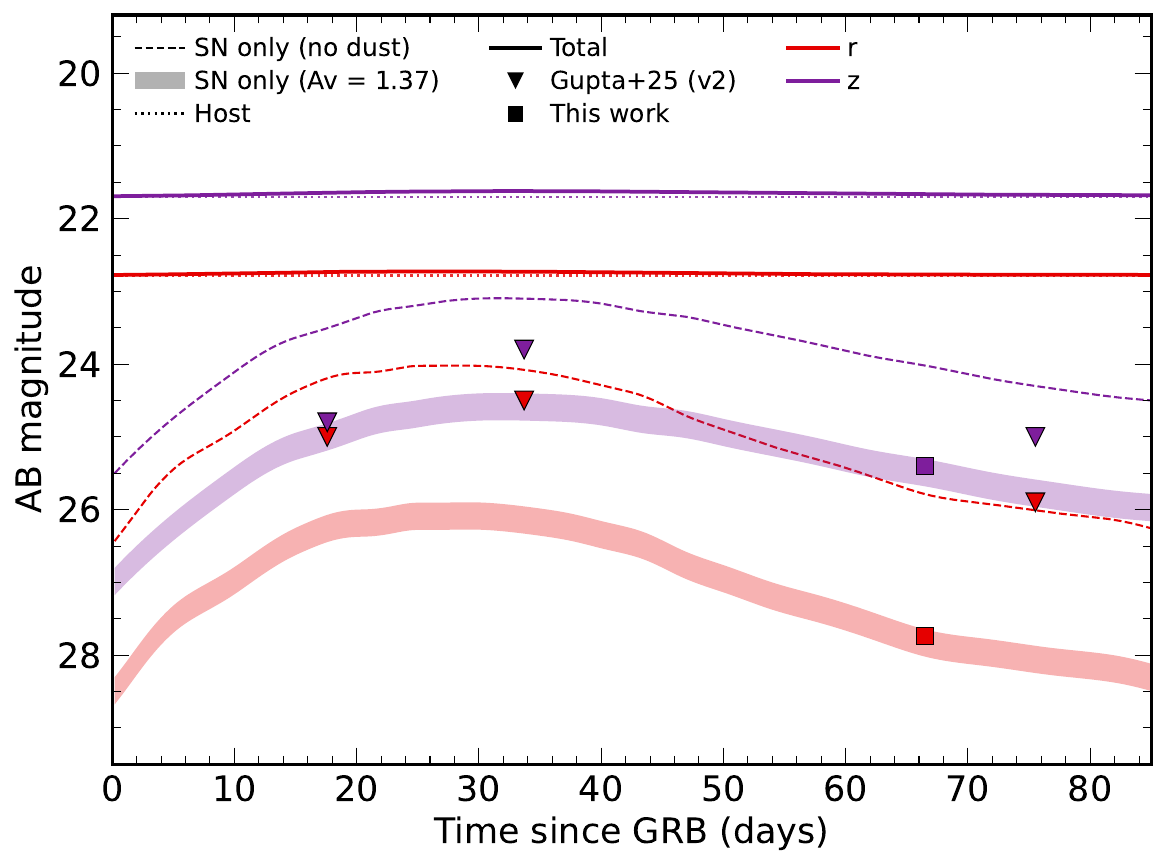} % \hsize
    \caption{Simulated optical $r$ (red) and $z$ (purple) light curves of the SN associated with GRB~240825A, modeled with SN~1998bw at $z = 0.659$, stretched by $s = 1.21$, scaled by $0.85$--$1.2$ in flux and reddened by $A^{\rm GRB}_V = 1.37$~mag (shaded bands). Dashed lines show the SN without host extinction, dotted lines the host galaxy, and solid lines the host plus SN. Triangles are $3\sigma$ upper limits from \citet{gupta2026a}. Squares are synthetic magnitudes of the SN~1998bw template matched to the host-subtracted JWST spectrum at 66.5~d.}
    \label{fig:sn98bw_lc}
\end{figure*}

We fix the host extinction towards the transient to $A^{\rm GRB}_{V}=1.37$~mag and adopt the \citet{fitzpatrick1999a} extinction law with $R_V=3.1$, evaluated at rest-frame wavelengths.
Dust attenuation of the stellar population is treated separately as a \prospector parameter. We place a Gaussian prior on the post-explosion template epoch, centered at 40 rest-frame days with a width of approximately 3 days. We also fit an additive white-noise term, combined in quadrature with the reported spectral uncertainties.

We first locate a high-posterior solution using differential evolution followed by local optimization. We then sample the posterior with the affine-invariant ensemble sampler \texttt{emcee} \citep{foreman-mackey2013a}, using 50 walkers initialized around the maximum-posterior point and 8000 steps, the first 4000 of which are discarded as burn-in.
The inferred intrinsic normalization of the transient relative to the SN~1998bw template in pixel 28 is $A_{{\rm SN},28}=0.42^{+0.02}_{-0.02}$, where the uncertainties correspond to the 16th and 84th posterior percentiles and include the covariance between the spatial amplitude, centroid, and width. This normalization describes only the transient contribution in pixel 28. It is neither an aperture-corrected total SN luminosity nor a bolometric luminosity ratio.
The preferred rest-frame template epoch is $t_{\rm SN}=33.05^{+0.58}_{-0.32}$~days. The compact component is centered at $\mu_{\rm SN}=28.55^{+0.06}_{-0.06}$, with a Gaussian width of $\sigma_{\rm SN}=0.95^{+0.06}_{-0.05}$~pixels. Under this profile, the transient flux assigned to pixel 30 is $0.37^{+0.07}_{-0.07}$ times that assigned to pixel 28. This ratio is conditional on the adopted spatial and spectral models and is not based on an independently calibrated NIRSpec PSF. Summing the SN component over pixels 26--35 for each posterior sample, we obtain a total normalization of $A_{\rm SN}=1.1\pm0.1$ relative to SN~1998bw.
For the galaxy component, the conditional posterior gives $\log(M_\star/M_\odot)=8.91^{+0.01}_{-0.01}$ and a star formation rate (SFR) of $\sim$1.8~$M_\odot\,{\rm yr}^{-1}$.

Figure~\ref{fig:joint_pixel28} shows the posterior decomposition for pixel 28. The model favors a non-zero transient contribution throughout the red part of the spectrum, including at observed wavelengths of $4$--$5\,\mu\mathrm{m}$.
The posterior predictions for all ten pixels are shown in Fig.~\ref{fig:joint_all_rows} and the SN fractions are listed in Table~\ref{tab:sn_fraction}. They peak at 43.9\% in pixel~28, and, integrated over all pixels, the SN contributes 13.1\% of the total observed flux.

The quoted statistical uncertainties are conditional on the adopted SN~1998bw spectral evolution and its infrared extrapolation, the fixed host extinction, and the wavelength-independent Gaussian spatial profile. The simple parametric galaxy model also does not reproduce the narrow nebular emission lines visible in the spectra. We therefore use the joint fit only to separate the broad host and transient continua, and do not interpret the narrow features or the detailed stellar population parameters. The systematic variation arising from alternative transient or galaxy templates is expected to dominate the formal sampling uncertainty.

\subsection{A supernova hidden from ground}
\label{ssec:sn_hidden}

\begin{figure*}[t]
    \sidecaption
    % \centering
    \includegraphics[width=12cm]{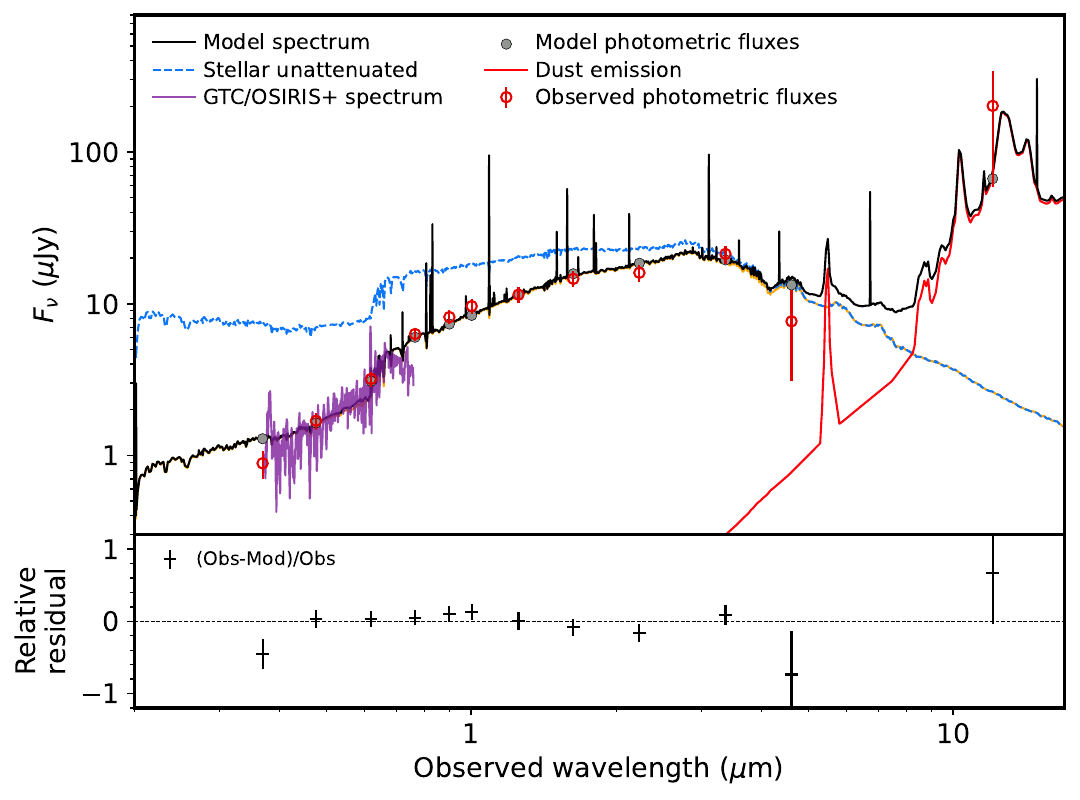} % \hsize
    \caption{Best-fit \cigale SED model ($\chi_\nu^2=0.97$) of the host galaxy of GRB~240825A. Observed fluxes are shown as red circles and model fluxes as black circles. The model spectrum is shown in black, the unattenuated stellar emission in blue, and the dust emission in red. The GTC/OSIRIS+ R1000B spectrum from \citet{gupta2026a}, smoothed with a Savitzky--Golay filter, is shown in purple. The lower panel shows the relative residuals of the photometric measurements.}
    \label{fig:cigale_sed_fit}
\end{figure*}

The JWST detection raises the question of why the SN was not detected in the deep optical and NIR follow-up observations of the GTC and the Large Binocular Telescope (LBT) reported by \citet{gupta2026a}. The combination of dust extinction, redshift, and host galaxy emission provides a natural explanation for their non-detection of the SN from the ground.
To better quantify these effects, we model the $r$ and $z$ band light curves a SN~1998bw-like event at $z=0.659$ using \texttt{SNCosmo} \citep{barbary2016a}. 
Both spectral decompositions presented in Sects.~\ref{ssec:empirical_host_sub} and \ref{sec:joint_host_transient} favor a SN~1998bw template at a phase of 54.85~days (33.1~days in the rest frame), earlier than the epoch of the JWST observations at 66.5~days after the trigger. We interpret this difference as a slower evolution than that of SN~1998bw and stretch the light curve in time by $s = 66.5/54.85 \simeq 1.21$. 
We scale the SN flux by 0.85--1.2 relative to SN~1998bw to encompass the normalizations obtained with the two decomposition methods and redden the model by $A^{\rm GRB}_V=1.37$~mag (Sect.~\ref{sec:joint_host_transient}).
At $z=0.659$, the observed $r$ and $z$ bands probe approximately the rest-frame $U$ and $V$ bands, respectively. The host extinction reduces the predicted SN peak by approximately 2.0~mag in $r$ and 1.4~mag in $z$. The simulated SN light curves (with and without dust) and the upper limits reported by \citet{gupta2026a} are shown in Fig.~\ref{fig:sn98bw_lc}.

At the first deep epoch of \citet{gupta2026a}, 17.6~days after the trigger (10.6~days in the rest frame), the SN was still rising due to its slower evolution relative to SN~1998bw. Its predicted magnitudes, $r=26.2$ and $z=24.9$, are approximately 1.2 and 0.1 mag fainter than their reported $3\sigma$ upper limits ($r>25.0$ and $z>24.8$). The SN detection was likely further complicated by the image subtraction. Although subtraction should remove most of the host contribution, reference images obtained $\lesssim$100~days after the burst may still contain SN light and affect the SN detection. 
At the second epoch, 33.7~days after the trigger (20.3~days in the rest frame), the SN was close to its peak, but the upper limits were shallower ($r>24.5$ and $z>23.8$), $\sim$1.6 and $\sim$0.7~mag brighter than the predicted SN magnitudes. At the last epoch, 75.5~days after the trigger (45.5~days in the rest frame), the SN had faded to $r=28.0$ and $z=25.7$, well below the upper limits of $r>25.9$ and $z>25.0$.
The SN therefore remained below the ground-based upper limits over the range of normalizations suggested by our analysis. However, these predictions should be interpreted with caution, as they are based on a single spectroscopic epoch, assuming that the SN evolved like the stretched and extrapolated SN~1998bw.
 
In contrast, our NIRSpec observations overcome these limitations by extending the wavelength coverage to the IR, where the dust extinction is lower, and by exploiting the angular resolution of JWST to increase the contrast between the SN and the host. At the epoch of the JWST observations, the SN contributes $\sim$44\% of the flux in pixel 28 (Table~\ref{tab:sn_fraction}) and $\sim$13\% in the spatially integrated 1D spectrum. In comparison, it would contribute only $\sim$1\% of the total flux in ground-based optical images at 66.5~days after the burst.

\subsection{Explosion site and local environment}
\label{ssec:local_env}
Relative astrometry between the X-shooter and archival HSC $r$-band images gives a projected offset of $\sim$3.27~kpc ($\sim$0.46$''$) between the GRB/SN position and the host galaxy center. The host is resolved in the HSC image, with an effective radius of $R_{\rm e}\sim4.1$~kpc measured with \texttt{SExtractor} \citep{bertin1996a}. This size is broadly consistent with those of star-forming field galaxies at this redshift \citep{vanderwel2014a}, but larger than is typical for long GRB hosts, which tend to be compact, dense galaxies \citep{kelly2014a,schneider2022a}. 
Although the physical offset exceeds the median values of $\sim$1.0--1.3~kpc reported for long GRBs \citep{blanchard2016a,lyman2017a}, it corresponds to a host-normalized offset of $\sim$$0.8\,R_{\rm e}$, only slightly above the median host-normalized offsets of $\sim$0.6 found for long GRBs \citep{blanchard2016a,lyman2017a}, and well below the median of $\sim$1.5 for short GRBs \citep{fong2013a}.

To investigate the local environment of the burst, we use the pixel-by-pixel JWST spectra. The joint host--transient decomposition places the SN at pixel 28.55 in the 2D spectrum, close to the position of 28.28 returned by \texttt{extract\_1d.source\_location} in the JWST pipeline.
The difference of 0.27~pixels corresponds to $\sim$0.03$''$ ($\sim$0.2~kpc) and is consistent with the observed excess in pixel 28 (Fig.~\ref{fig:jwst_spectrum_line_by_line}).
We estimate the local SFR from the \paalpha flux measured in each pixel spectrum, using the calibration of \citet{reddy2023a}. We do not apply a dust correction given that the \halpha line is likely contaminated by the broad SN features (Figs.~\ref{fig:jwst_spectrum_joint_fit_pix30_sub} and \ref{fig:joint_pixel28}), which prevents a reliable measurement of the \halpha/\paalpha ratio. In contrast, \paalpha\ is expected to be less affected by the SN signal and less sensitive to dust extinction. For $A_V = 1.5$\,mag, we expect a flux loss of $\approx$15\% at \paalpha, compared to $\approx$65\% for \halpha \citep{pei1992a}.
The SFRs measured in pixels 29 and 30 are similar, approximately $1~\Msun\,\mathrm{yr^{-1}}$ per pixel, but this value decreases to approximately $0.5~\Msun\,\mathrm{yr^{-1}}$ at the GRB/SN position.

The GRB/SN position is therefore slightly offset from the local peak of star formation along the slit. Although the short lifetimes of massive stars favor an association with recent star formation, they do not need to explode at its brightest peak. As discussed by \citet{christensen2008a}  \citep[see also][]{schady2017a}, GRB~980425/SN~1998bw occurred approximately 860~pc from a prominent Wolf--Rayet region, initially motivating a kicked (runaway) progenitor interpretation \citep{hammer2006a}. However, subsequent constraints on stellar ages favored formation at the explosion site, since travel from the nearby region would have required an unusually high velocity \citep{kruhler2017a}. Similarly, the offset observed for GRB~240825A could result from a progenitor kick, but could also reflect formation in a less luminous star-forming region. The current data do not allow us to distinguish between these scenarios.

\subsection{Host galaxy properties}
To characterize the large-scale host galaxy properties, we combine the broadband photometry of Table~\ref{tab:grb240825a_host_photometry} with the nebular emission lines detected in the X-shooter spectrum. In addition to the GRB afterglow, this spectrum shows several strong emission lines from the host galaxy \citep[Fig.~\ref{fig:xshooter_spec}; see also][]{wu2025a}, including \OIIab, \hgamma, \hbeta, \OIIIb, \halpha, \NIIa, and \SIIa. We model the atmospheric transmission for the source position and observing conditions using \textsc{SKYCALC} v2.0.9 \citep{noll2012a,jones2013a}\footnote{\url{https://www.eso.org/observing/etc/bin/gen/form?INS.MODE=swspectr+INS.NAME=SKYCALC}}, and measure the emission line fluxes by simultaneously fitting a Gaussian profile and the atmospheric transmission with the \textsc{lmfit} Python package \citep{newville2025a}. The measured fluxes are listed in Table~\ref{tab:host_em_line_flux}.
We note that the host photometry used in the SED fit was obtained 27.4 and 33.4~days after the burst, close to the predicted SN peak (Sect.~\ref{ssec:sn_hidden}). However, our simulations suggest that the SN contributes $\sim$5\% and $\sim$8\% of the host flux in the $r$ and $z$ bands at the SN peak ($\sim$30~days after the trigger, Fig.~\ref{fig:sn98bw_lc}). This contribution is comparable to the photometric and modeling uncertainties, and its impact on the derived host properties should be limited.

We model the host galaxy SED with \cigale \citep[Code Investigating GALaxy Emission;][]{boquien2019a}, fitting the broadband photometry together with the \halpha flux. We adopt the \citet{bruzual2003a} stellar population synthesis models with a \citet{chabrier2003a} initial mass function, allowing metallicities of $Z=0.008$ and 0.020, and include nebular emission with default parameters ($\log U=-2.0$, $Z_{\rm gas}=0.020$, and $n_{\rm e}=100~{\rm cm}^{-3}$). Dust attenuation is modeled with the modified \citet{calzetti2000a} law, with $E(B-V)_{\rm lines}$ allowed to vary between 0 and 1.5~mag and $E(B-V)_{\rm stars}=0.44\,E(B-V)_{\rm lines}$. We consider a delayed star formation history (SFH) with an optional ongoing burst of star formation, as observed in some GRB hosts \citep[e.g.,][]{corre2018b}. For the main stellar population, we allow the $e$-folding time to vary from 0.5 to 10~Gyr and impose a minimum age of 5~Gyr. The age of the recent burst can take values of 20, 50, or 100~Myr, and the burst mass fraction is explored from 0.0 to 0.5 in steps of 0.01.

The best-fit model, shown in Fig.~\ref{fig:cigale_sed_fit}, reproduces the photometric data well ($\chi_\nu^2=0.97$). The \halpha flux is also well reproduced, with a Bayesian estimate of $(24.8\pm2.7)\times10^{-17}~\mathrm{erg\,s^{-1}\,cm^{-2}}$. The posterior distribution of the burst mass fraction is consistent with zero, indicating that no recent burst is required. For comparison, we also overplot the host spectrum obtained with GTC/OSIRIS+ \citep[Optical System for Imaging and low-Intermediate-Resolution Integrated Spectroscopy;][]{cepa2000a} reported by \citet{gupta2026a}. The spectrum is arbitrarily scaled to match the best-fit model. Its continuum shows excellent agreement with our \cigale model and has a similar spectral slope.

From the \cigale fit, we infer a stellar mass of $\log(M_\star/M_\odot)=10.12\pm0.05$, an ${\rm SFR}=5.88\pm1.60~M_\odot\,{\rm yr}^{-1}$, and a nebular reddening of $E(B-V)_{\rm lines}=0.48\pm0.08$~mag, consistent with a moderately massive, actively star-forming, and dusty galaxy. Our stellar mass estimate is consistent with the analysis of \citet{gupta2026a}, who modeled the host with \prospector \citep{johnson2021a} and obtained $\log(M_\star/M_\odot)\simeq10.01$. However, our inferred SFR is higher than their estimate of $3.18~M_\odot\,{\rm yr}^{-1}$. This difference may partly arise from our inclusion of the \halpha flux and the $u_s$-band photometry, which provide additional constraints on recent star formation and the young stellar population. Our results imply a specific SFR (sSFR) of $\log({\rm sSFR}/{\rm yr}^{-1})\simeq-9.35$, consistent with the active star formation commonly observed in long GRB hosts \citep[e.g.,][]{savaglio2009a,svensson2010a,kruhler2015a,vergani2015a,palmerio2019a}.

We derive the gas-phase metallicity from the X-shooter emission lines, after correcting the line fluxes for the nebular extinction inferred from the \cigale SED fit ($E(B-V)_{\rm lines}=0.48$~mag, corresponding to $A_V\simeq1.5$~mag) and assuming a MW extinction curve \citep{pei1992a}. 
At $z=0.659$, the \NIIb line falls on a sky emission line and cannot be reliably measured. We infer its flux from \NIIa, which is measured in a clean spectral region (Table~\ref{tab:host_em_line_flux}). We adopt the theoretical ratio $F(\NIIb)/F(\NIIa)\simeq3$ \citep{storey2000a}.
Using the strong-line diagnostics of \citet{maiolino2008a}, we obtain $12+\log(\rm O/H) = 8.87 \pm 0.04$, corresponding to $Z = 1.51 \pm 0.14\,Z_\odot$ (assuming $12+\log(\rm O/H)_\odot = 8.69$; \citealt{asplund2009a}). This value is based on several line ratios (\OIIIb/\hbeta, \OIIab/\hbeta, \NIIb/\halpha, \OIIIb/\OIIab, and \OIIIb/\NIIb) and remains consistent for different dust extinction corrections tested. 

The metallicity is consistent with the mass--metallicity and fundamental metallicity relations of field star-forming galaxies at $z \sim 0.7$ \citep{mannucci2010a,mannucci2011a} but lies at the high end of the distribution of long GRB hosts. Compared with the BAT6 sample, analyzed with the same calibration \citep{japelj2016a,vergani2017a}, the host of GRB~240825A is more metal-rich than most GRB hosts of similar stellar mass and lies above the metallicity threshold for long GRB production derived by \citet{vergani2017a}, $Z_{\rm th}\simeq0.7\,Z_\odot$. However, several factors may reduce this tension. The metallicity is measured from the X-shooter slit spectrum, whereas the GRB lies at a projected offset of 3.3~kpc from the host center, in a region of lower star formation (Sect.~\ref{ssec:local_env}). A negative metallicity gradient could therefore imply a lower metallicity at the explosion site. Moreover, $Z_{\rm th}$ reflects a statistical suppression of the long GRB rate at high metallicity rather than a strict cutoff, and long GRBs with near-solar or super-solar host metallicities have been reported, particularly among dust-obscured events, which tend to occur in more massive and chemically enriched galaxies \citep[e.g.,][]{kruhler2011a,perley2013a,kruhler2015a,perley2016c}. GRB~240825A may therefore probe the metal-rich, dusty end of the long GRB host population.

%--------------------------------------------------------------------
\section{Discussion and conclusion}
\label{sec:Discussion_conclusions}
Our JWST observations reveal a point-like source at the position of GRB~240825A whose emission shows broad features similar to those of SN~1998bw and SN~2023lcr.
The identification of a SN Ic-BL through both the empirical host subtraction and the joint host-transient decomposition confirms that GRB~240825A originated from a massive-star.

The SN detection resolves the progenitor ambiguity reported by \citet{gupta2026a}. Their analysis yielded conflicting indications about the nature of the progenitor. 
The relatively short ($T_{90}\simeq4$~s), hard prompt episode ($E_p \simeq 405$~keV) and short variability timescale ($\sim$14~ms) suggested a possible merger origin \citep{gupta2026a}. In addition, the presence of a soft tail does not exclude this scenario, as extended emission has also been observed in some short GRBs \citep[e.g.,][]{barthelmy2005b,norris2006a}. 
On the other hand, the burst energetics, its position on the Amati relation, and several classification diagnostics favored a collapsar origin \citep{gupta2026a}. 
Finally, despite deep ground follow-up observations, no associated SN was detected, leaving the progenitor uncertain. By revealing the SN, the JWST spectrum provides direct evidence for the massive-star origin of GRB~240825A. The inferred dusty line of sight extinction ($A_V^{\rm GRB}=1.37\pm0.08$~mag), and the host contamination provide a natural explanation for the non-detection of the SN in ground-based observations. In turn, these observations demonstrate the ability of JWST to robustly isolate SN events even along dusty lines of sight and situated within bright host galaxies. Such capabilities may be crucial in the future to determine progenitors. Indeed, while there are ultimately limits to the ability of JWST to probe the dustiest environments (e.g., the progenitor of GRB~250702B with $A_V >10$ remains ambiguous \citep{levan2025c,gompertz2026a,sears2026a} we might expect to recover a SN, or place meaningful limits, in a substantial fraction of future cases. 

GRB~200826A, with a rest-frame duration of $\lesssim$1~s and a relatively soft prompt emission, was a short-duration GRB with an associated SN \citep{ahumada2021a,rossi2022b}, showing that a short duration alone does not necessarily imply a merger origin.
GRB~240825A extends this to a burst whose prompt emission is more typical of short GRBs. With a rest-frame duration of $\sim$2.4~s and a rest-frame peak energy of $E_{\rm p,z} \simeq 733$~keV (\textit{Fermi}/GBM), it lies within the short-GRB population in the $E_{\rm p,z}$--duration plane \citep{ahumada2021a}, well separated from the softer GRB~200826A.
In contrast, the KNe associated with the long-duration GRBs~211211A and 230307A show that a long duration does not necessarily imply a massive-star progenitor \citep[e.g.,][]{rastinejad2022a,troja2022a,yang2022a,gompertz2023a,levan2023b}. Spectroscopic identification of the accompanying transient is, therefore, essential for building GRBs samples with robust progenitor classification. Such samples are crucial for determining the relative rates and comparing the properties of the different progenitor populations, studying the evolution of GRB/SNe properties with redshift, and constraining the contribution of compact object mergers to the production of heavy elements via $r$-process.

The host galaxy properties also make this association intriguing. It is a dusty, moderately massive and actively star-forming galaxy with $\log(M_\star/M_\odot)=10.12\pm0.05$, ${\rm SFR}=5.88\pm1.60~M_\odot\,{\rm yr}^{-1}$, and a gas-phase metallicity of $12+\log({\rm O/H})=8.87\pm0.04$ (i.e., $Z = 1.51 \pm 0.14\,Z_\odot$). This places it at the metal-rich end of the collapsar GRB host population \citep{japelj2016a,vergani2017a}, rarely observed among this population. The explosion is located within its effective radius, at a projected offset of 3.3 kpc from the galaxy center, in a region with ongoing star formation, although at a lower level than in the neighboring regions traced by \paalpha along the slit or the SFR inferred from the SED fitting. 

These results show how dust extinction and host contamination can hide a SN, and potentially affect our GRB progenitor classification. The spatial resolution of JWST allows us to isolate the explosion site and increase the contrast against the host, revealing a SN that contributes only $\sim$10--15\% of the spatially integrated NIRSpec flux and would contribute only $\sim$1\% of the total flux in ground-based optical images at the JWST epoch (66.5 days after the trigger). Combined with its infrared sensitivity, it shows the unique capability of JWST to identify a population of dust-obscured SNe in dusty, enriched environments that were previously hard to assess and poorly explored.

%--------------------------------------------------------------------
\begin{acknowledgements}
    BS, VB, ELF, and SDV acknowledge the support of the French Agence Nationale de la Recherche (ANR), under grant ANR-23-CE31-0011 (project PEGaSUS).
    NS is supported by the Kavli Foundation.
    AR acknowledges financial support from PRIN-MIUR 2017 (grant 20179ZF5KS).
    DH acknowledges support from NSF PAARE.
    POB and NRT acknowledge support from STFC (NRT under grant UKRI1200).
    RS acknowledges the contribution by the Italian Space Agency, contract ASI/INAF n.I/004/11/6.
    The Cosmic Dawn Center (DAWN) is funded by the Danish National Research Foundation under grant DNRF140.
    DBM and DW acknowledge funding by the European Union (ERC, HEAVYMETAL, 101071865). Views and opinions expressed are, however, those of the authors only and do not necessarily reflect those of the European Union or the European Research Council. Neither the European Union nor the granting authority can be held responsible for them. \\

    This work is based in part on observations made with the NASA/ESA/CSA James Webb Space Telescope. The data were obtained from the Mikulski Archive for Space Telescopes at the Space Telescope Science Institute, which is operated by the Association of Universities for Research in Astronomy, Inc., under NASA contract NAS 5-03127 for JWST. These observations are associated with program \#6133. \\

    Based on data from the GTC Archive at CAB (CSIC -INTA). The GTC Archive is part of the Spanish Virtual Observatory project funded by MCIN/AEI/10.13039/501100011033 through grant PID2023-146210NB-I00. \\
    
    This paper is based on data collected at the Subaru Telescope and retrieved from the HSC data archive system, which is operated by the Subaru Telescope and Astronomy Data Center (ADC) at NAOJ. Data analysis was in part carried out with the cooperation of Center for Computational Astrophysics (CfCA), NAOJ. We are honored and grateful for the opportunity of observing the Universe from Maunakea, which has the cultural, historical and natural significance in Hawaii. \\

    The Pan-STARRS1 Surveys (PS1) and the PS1 public science archive have been made possible through contributions by the Institute for Astronomy, the University of Hawaii, the Pan-STARRS Project Office, the Max-Planck Society and its participating institutes, the Max Planck Institute for Astronomy, Heidelberg and the Max Planck Institute for Extraterrestrial Physics, Garching, The Johns Hopkins University, Durham University, the University of Edinburgh, the Queen's University Belfast, the Harvard-Smithsonian Center for Astrophysics, the Las Cumbres Observatory Global Telescope Network Incorporated, the National Central University of Taiwan, the Space Telescope Science Institute, the National Aeronautics and Space Administration under Grant No. NNX08AR22G issued through the Planetary Science Division of the NASA Science Mission Directorate, the National Science Foundation Grant No. AST-1238877, the University of Maryland, Eotvos Lorand University (ELTE), the Los Alamos National Laboratory, and the Gordon and Betty Moore Foundation. \\

    SDSS is managed by the Astrophysical Research Consortium for the Participating Institutions of the SDSS Collaboration, including the Carnegie Institution for Science, Chilean National Time Allocation Committee (CNTAC) ratified researchers, Caltech, the Gotham Participation Group, Harvard University, Heidelberg University, The Flatiron Institute, The Johns Hopkins University, L'Ecole polytechnique f\'{e}d\'{e}rale de Lausanne (EPFL), Leibniz-Institut f\"{u}r Astrophysik Potsdam (AIP), Max-Planck-Institut f\"{u}r Astronomie (MPIA Heidelberg), Max-Planck-Institut f\"{u}r Extraterrestrische Physik (MPE), Nanjing University, National Astronomical Observatories of China (NAOC), New Mexico State University, The Ohio State University, Pennsylvania State University, Smithsonian Astrophysical Observatory, Space Telescope Science Institute (STScI), the Stellar Astrophysics Participation Group, Universidad Nacional Aut\'{o}noma de M\'{e}xico, University of Arizona, University of Colorado Boulder, University of Illinois at Urbana-Champaign, University of Toronto, University of Utah, University of Virginia, Yale University, and Yunnan University. \\ 

    This publication makes use of data products from the Wide-field Infrared Survey Explorer, which is a joint project of the University of California, Los Angeles, and the Jet Propulsion Laboratory/California Institute of Technology, funded by the National Aeronautics and Space Administration.

\end{acknowledgements}

%--------------------------------------------------------------------
\bibliographystyle{aa}
\bibliography{ref}

%--------------------------------------------------------------------
\begin{appendix}
\onecolumn
\section{Additional materials}
\label{app}

\begin{table*}[!htbp]
\centering
\renewcommand{\arraystretch}{1.35}
\caption{Best-fit parameters of the power-law model for the NIR-to-X-ray afterglow SED of GRB~240825A derived from the spectroscopic data set.}
\label{tab:240825Afitresults}
\label{tab:240825Afitsed}
\resizebox{\textwidth}{!}{
\begin{tabular}{ccccccccccc}
\hline\hline
\makecell{Extinction \\ curve} &
$\beta_\mathrm{OX}$ &
\makecell{$E(B-V)$ \\ mag} &
$R_\mathrm{V}$ &
\makecell{$A_\mathrm{V}$ \\ mag} &
$c_1$ &
$c_2$ &
$c_3$ &
$c_4$ &
\makecell{$N_{\rm H,X}$ \\ $10^{22}~\text{cm}^{-2}$} &
\makecell{$\chi^2_\mathrm{red}$ \\ (d.o.f)} \\
\hline
FM &
$0.76_{-0.02}^{+0.01}$ &
$0.37_{-0.02}^{+0.02}$ &
$3.72_{-0.11}^{+0.11}$ &
$1.37_{-0.08}^{+0.08}$ &
$1.82_{-1.56}^{+1.78}$ &
$0.42_{-0.52}^{+0.46}$ &
$1.01_{-0.58}^{+0.67}$ &
$0.41_{-0.10}^{+0.15}$ &
$0.79_{-0.10}^{+0.08}$ &
$0.77\,(1289)$ \\
\hline
SMC &
$0.78_{-0.01}^{+0.01}$ &
$-$ &
$-$ &
$1.17_{-0.01}^{+0.01}$ &
$-$ &
$-$ &
$-$ &
$-$ &
$0.78_{-0.08}^{+0.09}$ &
$0.88\,(1292)$ \\
LMC &
$0.79_{-0.01}^{+0.01}$ &
$-$ &
$-$ &
$1.26_{-0.02}^{+0.02}$ &
$-$ &
$-$ &
$-$ &
$-$ &
$0.81_{-0.08}^{+0.10}$ &
$0.84\,(1292)$ \\
MW &
$0.79_{-0.01}^{+0.01}$ &
$-$ &
$-$ &
$1.31_{-0.02}^{+0.02}$ &
$-$ &
$-$ &
$-$ &
$-$ &
$0.80_{-0.08}^{+0.09}$ &
$0.94\,(1292)$ \\
\hline
\end{tabular}
}\par\smallskip
\begin{minipage}{\textwidth}
\small \textbf{Notes:} Column (1): extinction curve used in the fit. Column (2): NIR-to-X-ray spectral slope $\beta_{\mathrm{OX}}$. The subscript follows the usual optical-to-X-ray notation, although the fit also includes the NIR data. Column (3): host galaxy reddening $E(B-V)$. Column (4): total-to-selective extinction ratio $R_{\mathrm{V}}$. Column (5): visual extinction $A_{\mathrm{V}}$. Columns (6--9): UV linear intercept $c_1$, UV slope $c_2$, bump strength $c_3$, and far-UV curvature $c_4$. Column (10): equivalent hydrogen column density $N_{\mathrm{H,X}}$. Column (11): reduced $\chi^2$ and number of degrees of freedom (d.o.f.). Uncertainties are given at the $1\sigma$ level.
\end{minipage}
\end{table*}

\begin{table*}[!htbp]
\centering
\renewcommand{\arraystretch}{1.0}
\begin{minipage}[t]{0.48\textwidth}
\vspace{0pt}
    \caption{Emission line fluxes of the GRB~240825A host galaxy.}
    \centering
    \begin{tabular}{cc}
         \hline
         Emission line & Flux \\
        \hline
        \OIIab  & $8.7 \pm 0.6$ \\
        \hgamma & $1.1 \pm 0.2$ \\
        \hbeta  & $4.2 \pm 0.2$ \\
        \OIIIb  & $3.8 \pm 0.4$ \\
        \halpha & $27.9 \pm 2.6$ \\
        \NIIa   & $2.7 \pm 0.6$ \\
        \SIIa   & $4.3 \pm 1.5$ \\
        \hline
    \end{tabular}
    \label{tab:host_em_line_flux}
\tablefoot{Fluxes are given in units of $10^{-17}~\mathrm{erg\,s^{-1}\,cm^{-2}}$. They are corrected for Galactic extinction ($A_{\rm V}^{\rm Gal} = 0.17$~mag) and for slit losses (factor of 2.15, based on broadband photometry obtained prior to the spectrum), but not for host-galaxy extinction.}
\end{minipage}\hfill
\begin{minipage}[t]{0.48\textwidth}
\vspace{0pt}
\centering
\caption{Photometry of the GRB~240825A host galaxy.}
\label{tab:grb240825a_host_photometry}
\begin{tabular}{lcc}
\hline
Telescope/Instrument & Filter & Magnitude (AB) \\
\hline
GTC/HiPERCAM & $u_s$ & $24.29 \pm 0.20$ \\
GTC/HiPERCAM & $g_s$ & $23.54 \pm 0.08$ \\
GTC/HiPERCAM & $r_s$ & $22.78 \pm 0.05$ \\
GTC/HiPERCAM & $i_s$ & $22.01 \pm 0.03$ \\
GTC/HiPERCAM & $z_s$ & $21.70 \pm 0.04$ \\
Subaru/HSC & $y$ & $21.51 \pm 0.05$ \\
GTC/EMIR  & $J$ & $21.29 \pm 0.08$ \\
GTC/EMIR & $H$ & $21.02 \pm 0.07$ \\
GTC/EMIR & $K$ & $20.91 \pm 0.08$ \\
WISE & $W1$ & $20.58 \pm 0.10$ \\
WISE & $W2$ & $21.69 \pm 0.64$ \\
WISE & $W3$ & $18.14 \pm 0.76$ \\

\hline
\end{tabular}
\tablefoot{Magnitudes are in the AB system and are not corrected for Galactic extinction.}
\end{minipage}
\end{table*}

\begin{table*}[!htbp]
\centering
\renewcommand{\arraystretch}{1.1}

\begin{minipage}[t]{0.49\textwidth}
\vspace{0pt}
\centering
\caption{SN contribution to the observed flux in each pixel spectrum.}
\label{tab:sn_fraction}
\begin{tabular}{ccc}
\hline
Pixel & Median SN fraction (\%) & 90\% credible interval \\
\hline
26     & 19.2 & 10.2--30.7 \\
27     & 40.5 & 32.9--47.3 \\
28     & 43.9 & 41.1--47.0 \\
29     & 25.8 & 23.8--27.9 \\
30     & 8.0  & 5.7--10.4 \\
31     & 1.2  & 0.5--2.3 \\
32     & 0.07 & 0.01--0.22 \\
33--35  & $<$0.01 & \\
\hline
All pixels & 13.1 & 11.9--14.3 \\
\hline
\end{tabular}
\tablefoot{The SN fraction is the fraction of the total observed flux attributed to the SN component derived from the joint host--transient spectral decomposition.}
\end{minipage}

\end{table*}

\begin{figure*}[!htbp]
    \sidecaption
    \includegraphics[width=12cm]{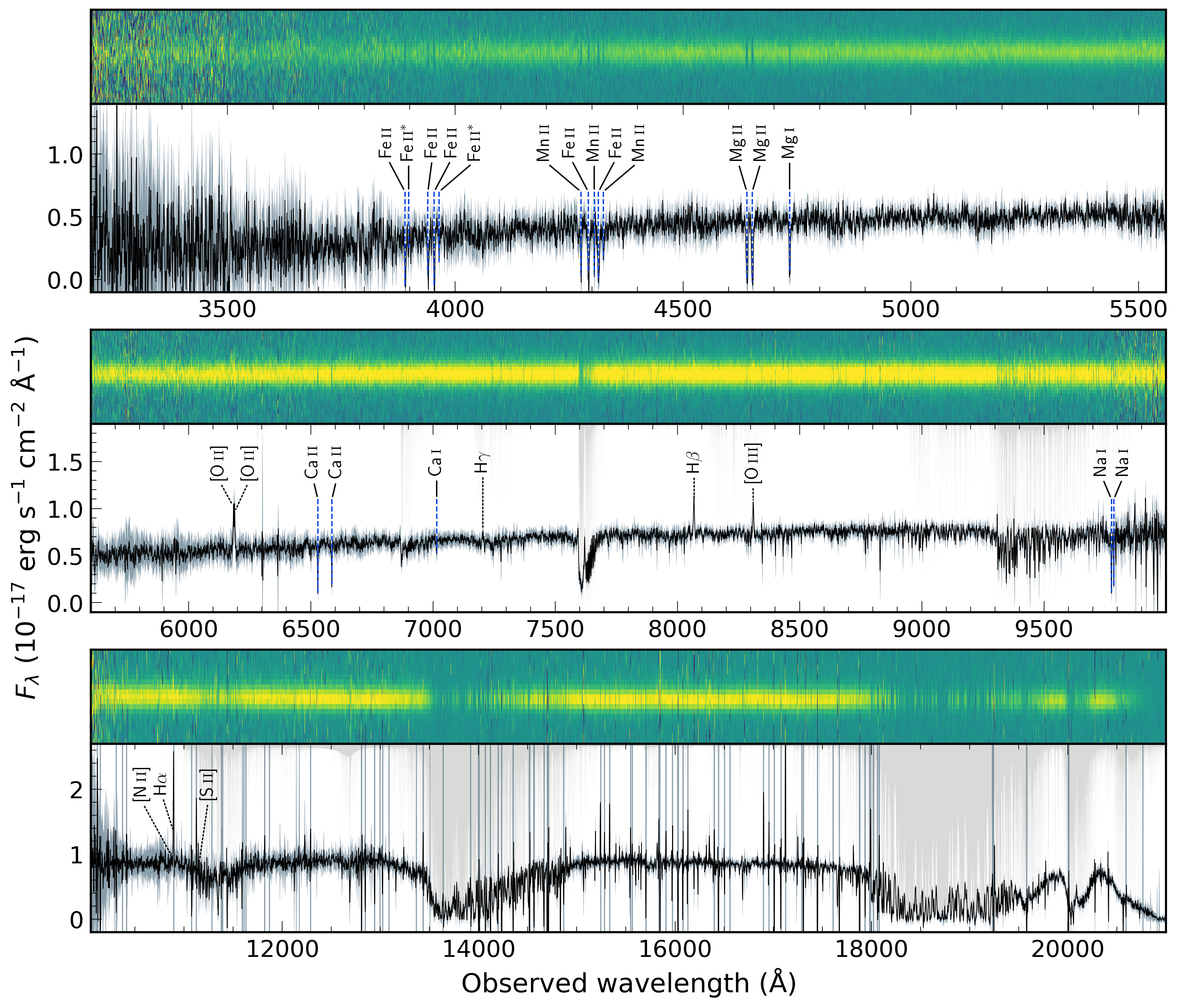}
    \caption{VLT/X-shooter spectrum of GRB~240825A ($z = 0.659$) obtained 11.35~hr after the burst. The UVB, VIS, and NIR arms are shown in the top, middle, and bottom panels, respectively. For each arm, the upper subpanel displays the 2D spectrum and the lower subpanel the extracted 1D spectrum (black). Absorption features from the GRB host galaxy and emission lines are marked by blue dashed and black dotted vertical lines, respectively. For clarity, the 1D spectra are smoothed with a Savitzky--Golay filter. The spectrum uncertainties are shown in blue-gray around the data and regions of telluric absorption are shaded in gray.}
    \label{fig:xshooter_spec}
\end{figure*}

\begin{figure*}[!htbp]
    \sidecaption
    \includegraphics[width=12cm]{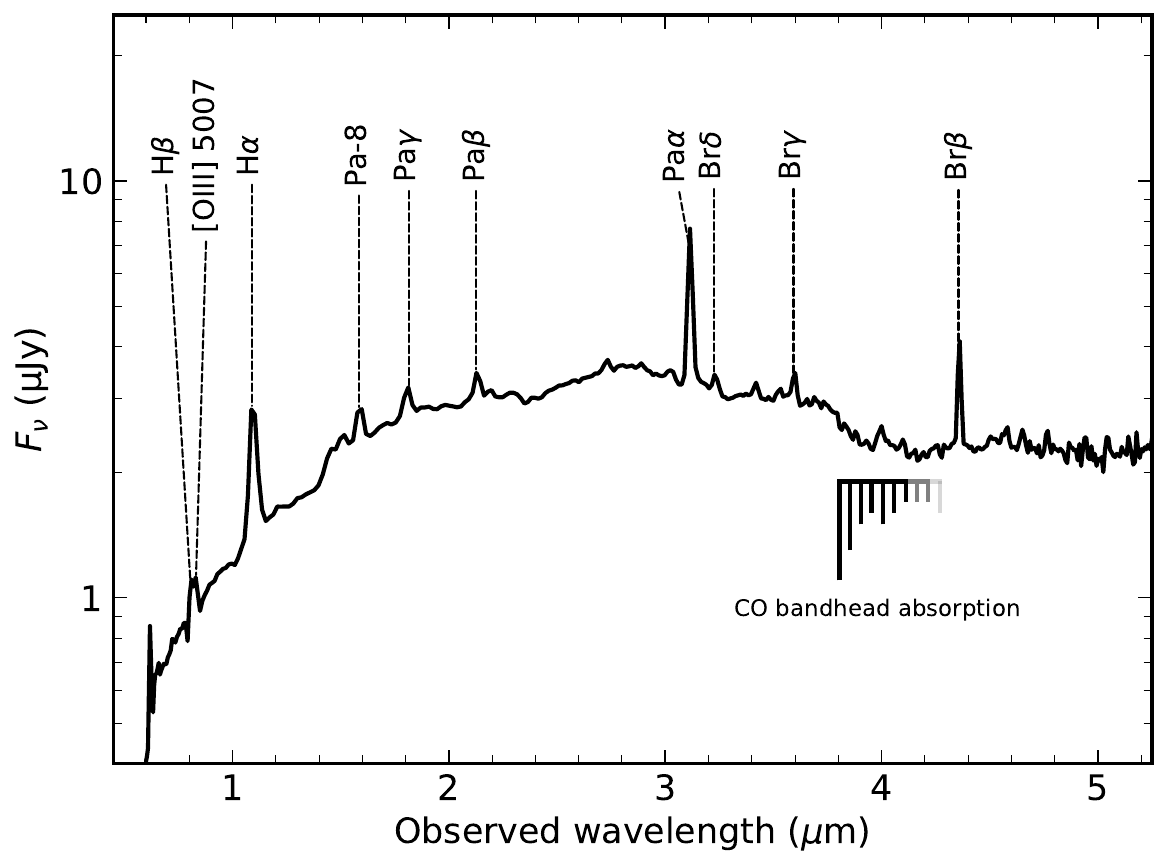}
    \caption{JWST/NIRSpec spectrum of GRB~240825A 66.5~days (40.1~days in the rest frame) after the trigger. The main host-galaxy emission lines, marked by vertical dashed lines, include \hbeta, \OIIIb, \halpha, and several Paschen and Brackett recombination lines. The expected wavelengths of individual bandhead components (CO 2–0 through CO 11–9) are indicated in progressively lighter shades.}
    \label{fig:jwst_spectrum_1D_lines}
\end{figure*}

\begin{figure*}[!htbp]
 \centering
 \includegraphics[width=17cm]{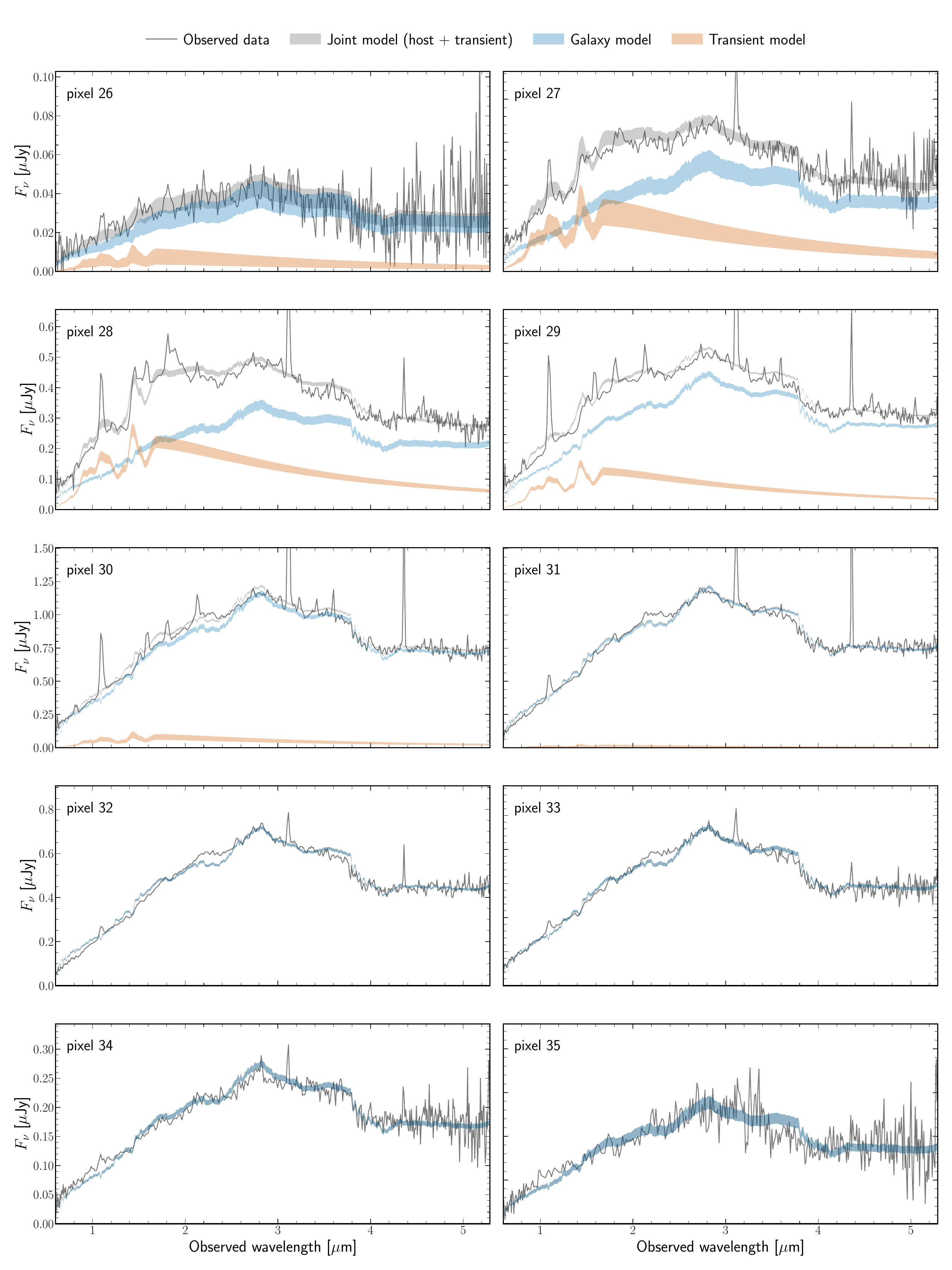}
 \caption{Posterior predictions from the simultaneous fit to spatial pixels 26--35. Gray, blue, and orange bands show the 90\% posterior intervals for the total model, the galaxy, and the compact SN~1998bw-like component, respectively. The measured spectra and their reported uncertainties are shown in gray.}
 \label{fig:joint_all_rows}
\end{figure*}

\end{appendix}
\end{document}